\documentclass[11pt]{article}

\usepackage{fullpage}
\usepackage[T1]{fontenc}
\usepackage[usenames,dvipsnames]{xcolor}
\usepackage{stmaryrd}

\usepackage{mathpazo}
\usepackage{xfrac}
\usepackage{bm}
\usepackage{todonotes}
\usepackage{scrextend}
\usepackage{enumitem}

\RequirePackage[colorlinks=true,pagebackref=true]{hyperref}
\hypersetup{
  linkcolor=[rgb]{0.1,0.5,0.5},
  citecolor=[rgb]{0.5, 0.1, 0.5},
  urlcolor=[rgb]{0.5, 0.1, 0.5}
}

\usepackage{atmacros}
\usepackage{amsthm}
\usepackage{thmtools}
\usepackage{thm-restate}

\numberwithin{equation}{section}
\declaretheoremstyle[bodyfont=\it,qed=\qedsymbol]{noproofstyle}

\declaretheorem[numberlike=equation]{observation}
\declaretheorem[numberlike=equation,style=noproofstyle,name=Observation]{observationwp}
\declaretheorem[name=Observation,numbered=no]{observation*}

\declaretheorem[numberlike=equation]{theorem}
\declaretheorem[numberlike=equation,style=noproofstyle,name=Theorem]{theoremwp}
\declaretheorem[name=Theorem,numbered=no]{theorem*}

\declaretheorem[numberlike=equation]{lemma}
\declaretheorem[name=Lemma,numbered=no]{lemma*}
\declaretheorem[numberlike=equation,style=noproofstyle,name=Lemma]{lemmawp}

\declaretheorem[name=Corollary,numbered=no]{corollary*}

\declaretheorem[numberlike=equation]{proposition}
\declaretheorem[name=Proposition,numbered=no]{proposition*}
\declaretheorem[numberlike=equation,style=noproofstyle,name=Proposition]{propositionwp}

\declaretheorem[numberlike=equation]{claim}
\declaretheorem[name=Claim,numbered=no]{claim*}

\declaretheorem[name=Conjecture,numbered=no]{conjecture*}

\declaretheorem[name=Question,numbered=no]{question*}

\declaretheoremstyle[bodyfont=\it,qed=$\lozenge$]{defstyle} 

\declaretheorem[numberlike=equation,style=defstyle]{definition}
\declaretheorem[unnumbered,name=Definition,style=defstyle]{definition*}

\declaretheorem[unnumbered,name=Example,style=defstyle]{example*}

\declaretheorem[unnumbered,name=Notation=defstyle]{notation*}

\declaretheorem[unnumbered,name=Construction,style=defstyle]{construction*}

\declaretheorem[numberlike=equation,style=defstyle]{remark}
\declaretheorem[unnumbered,name=Remark,style=defstyle]{remark*}

\newif\ifdraft
\drafttrue 

\ifdraft

\newcommand{\PCnote}[1]{\textcolor{BrickRed}{\guillemotleft PC: #1 \guillemotright}}
\newcommand{\ATnote}[1]{\textcolor{OliveGreen}{\guillemotleft AT: #1 \guillemotright}}
\newcommand{\pending}[1]{#1 \vspace{1em}}
\else

\newcommand{\PCnote}[1]{}
\newcommand{\ATnote}[1]{}
\newcommand{\pending}[1]{}
\fi

\newcommand{\ignore}[1]{}

\newcommand{\inparen}[1]{\left( #1 \right)}

\newcommand{\inbrace}[1]{\left\{ #1 \right\}}

\newcommand{\inangle}[1]{\left< #1 \right>}

\newcommand{\set}[1]{\inbrace{#1}}

\newcommand{\abs}[1]{\left| #1 \right|}

\newcommand{\ceil}[1]{\left\lceil #1 \right\rceil}

\newcommand{\coeff}{\operatorname{coeff}}
\newcommand{\size}{\operatorname{size}}

\newcommand{\cvector}{\overline{\operatorname{coeff}}}

\newcommand{\byPoly}[1]{#1_{\mathsf{/poly}}}

\newcommand{\CH}{\mathsf{CH}}

\newcommand{\NP}{\mathsf{NP}}
\def\P{\mathsf{P}}
\newcommand{\PSPACE}{\mathsf{PSPACE}}
\newcommand{\SPACE}{\mathsf{SPACE}}

\newcommand{\VP}{\mathsf{VP}}

\newcommand{\VPSPACE}{\mathsf{VPSPACE}}
\newcommand{\VNP}{\mathsf{VNP}}

\newcommand{\calc}{\mathcal{C}}
\newcommand{\cald}{\mathcal{D}}
\newcommand{\calg}{\mathcal{G}}
\newcommand{\calh}{\mathcal{H}}

\newcommand{\calp}{\mathcal{P}}
\newcommand{\calu}{\mathcal{U}}

\newcommand{\C}{\mathbb{C}}

\def\phi{\varphi}   
\def\epsilon{\varepsilon}   
\newcommand{\NC}{\mathsf{NC}}
\newcommand{\TC}{\mathsf{TC}}

\newcommand{\VPPROJ}{\mathsf{VPPROJ}}

\newcommand{\proj}{\operatorname{fix}}

\newcommand{\summation}{\mathtt{sum}}
\newcommand{\production}{\mathtt{prod}}

\usepackage{nth}
\usepackage{intcalc}
\usepackage{etoolbox}
\usepackage{xstring}

\usepackage{ifpdf}
\ifpdf
\else
\usepackage[quadpoints=false]{hypdvips}
\fi

\newcommand{\ECCCurl}[2]{http://eccc.hpi-web.de/report/\ifnumcomp{#1}{>}{93}{19}{20}#1/#2/}

\newcommand{\shortECCC}[2]{\texttt{\href{http://eccc.hpi-web.de/report/\ifnumcomp{#1}{>}{93}{19}{20}#1/#2/}{eccc:TR#1-#2}}}

\newcommand{\parseECCC}[1]{
\StrSubstitute{#1}{TR}{}[\tmpstring]%
\IfSubStr{\tmpstring}{/}{ 
\StrBefore{\tmpstring}{/}[\ecccyear]%
\StrBehind{\tmpstring}{/}[\ecccreport]%
}{
\StrBefore{\tmpstring}{-}[\ecccyear]%
\StrBehind{\tmpstring}{-}[\ecccreport]%
}%
\shortECCC{\ecccyear}{\ecccreport}}

\title{Lower Bounds from Succinct Hitting Sets}
\author{
 	Prerona Chatterjee \thanks{Department of Mathematics, IIT Madras. Parts of this work were supported by the Azrieli International Postdoctoral Fellowship, the Israel Science Foundation (grant number 514/20), the Len Blavatnik and the Blavatnik Family foundation, and a fellowship of the DAE, India for TIFR Mumbai. Email: \texttt{prerona.ch@gmail.com}.}
 	\and
 	Anamay Tengse \thanks{School of Computer Sciences, NISER, Bhubaneswar. Parts of this work were supported by the Israel Science Foundation (grant No. 716/20 and grant No. 843/23), and a fellowship of the DAE, India for TIFR Mumbai. Email: \texttt{anamay.tengse@gmail.com}.}
}

\date{}

\begin{document}

\maketitle
 
\pagenumbering{gobble}

\begin{abstract}

We investigate the consequences of the existence of ``efficiently describable'' hitting sets for polynomial sized algebraic circuit ($\VP$), in particular, \emph{\textsf{VP}-succinct hitting sets}.
Existence of such hitting sets is known to be equivalent to a ``natural-proofs-barrier'' towards algebraic circuit lower bounds, from the works that introduced this concept~\cite{FSV18,GKSS17}.
We show that the existence of \textsf{VP}-succinct hitting sets for \textsf{VP} would either imply that $\VP \neq \VNP$, or yield a fairly strong lower bound against $\TC^0$ circuits, assuming the Generalized Riemann Hypothesis (GRH).

This result is a consequence of showing that designing efficiently describable ($\VP$-explicit) hitting set generators for a class $\mathcal{C}$, is essentially the same as proving a separation between $\mathcal{C}$ and $\VPSPACE$: the algebraic analogue of \textsf{PSPACE}.
More formally, we prove an upper bound on \emph{equations} for polynomial sized algebraic circuits (\textsf{VP}), in terms of \textsf{VPSPACE}.

Using the same upper bound, we also show that even \emph{sub-polynomially explicit hitting sets} for \textsf{VP} --- much weaker than \textsf{VP}-succinct hitting sets that are almost polylog-explicit --- would imply that either $\VP \neq \VNP$ or that $\P \neq \PSPACE$.
This motivates us to define the concept of \emph{cryptographic hitting sets}, which we believe is interesting on its own.
  
\end{abstract}

\newpage

\pagenumbering{arabic}
\setcounter{page}{1}

\section{Introduction}\label{intro}

Proving lower bounds against boolean circuits is extremely hard. This is not news to anyone in the area, and perhaps, coming to terms with this reality is a rite of passage for all complexity theorists.
More concretely, the famous work of Razborov and Rudich~\cite{RR97} rules out most ``natural'' strategies for proving lower bounds against any boolean circuit class that is rich enough to do strong-enough cryptography.

Every boolean function can be seen as a multilinear polynomial, and it is therefore quite natural to try and understand the complexity of computing these polynomials \emph{syntactically} (using sum and product gates).
Computing a function syntactically can only be a harder task, since any such implementation must compute the same function.
This suggests that proving algebraic circuit lower bounds should only be easier.

Take the case of the $\mathtt{\#SAT}$ function, whose syntactic counterpart is the Permanent polynomial~\cite{V79b}.
The latter captures the algebraic counterpart of $\NP$, called $\VNP$~\cite{V79a}, and it is known that showing $\VP = \VNP$ (``algebraic $\P = \NP$'') would immediately give efficient boolean circuits for $\mathtt{SAT}$, barring some unlikely issues due to field constants~\cite{B00}.
It can further be shown that any algebraic circuit computing a degree $d$ polynomial does not benefit much from having intermediate computations of degree more than $d$~\cite{S73}.
Algebraic circuit complexity therefore focusses on studying the complexity of syntactically computing \emph{low-degree} polynomials using algebraic operations of sums and products as gates.
The reader may refer to some comprehensive surveys~\cite{SY10,CKW11,S15} for an overview of the area.

Despite this supposed advantage towards proving hardness, almost all interesting algebraic circuit classes have been doing a pretty good impression of their boolean counterparts as they continue to resist some promising attacks (like shifted partials: \cite{AV08,GKKS14,FLMS15}, set-multilinearization: \cite{BDS24,CKSS24}).
As a result, the one-line summary of the best known lower bounds does not read much better than the boolean world: $\Omega(n \log n)$ for circuits~\cite{BS83,S97b}, $\Omega(n^2)$ for ABPs~\cite{CKSV22}, $\Omega(n^2)$ for formulas~\cite{K85,CKSV22}.

\subsection*{Algebraic Natural Proofs}

This state of affairs has inspired studies in search of a formal barrier towards proving strong algebraic circuit lower bounds using the conventional methods, along the lines of the aforementioned work of Razborov and Rudich~\cite{RR97}.
The analogous notion that captures almost all known algebraic circuit lower bounds, is the framework of \emph{algebraic(ally) natural proofs}.
This framework was jointly proposed in the works of Forbes, Shpilka and Volk~\cite{FSV18}, and Grochow, Kumar, Saks and Saraf~\cite{GKSS17} achieves this well\footnote{These build on the works of Aaronson and Drucker~\cite{AD08}, and Grochow~\cite{G15}.}.
This framework is obtained essentially by ``algebraizing'' the framework of natural proofs from \cite{RR97}.

Intuitively, say we wish to prove a lower bound of $s$ on the size of any circuit computing a polynomial $f$.
This can be viewed as a task of proving non-membership in a set: the set $\mathcal{C}$ that contains all polynomials computable with size less than $s$, and we want to show $f \not\in \mathcal{C}$.
What would help here is a polynomial $D$ that can certify this fact.
In particular, if $D$ were to evaluate to zero on all members of $\mathcal{C}$, and to some nonzero value on $f$, then $D$ would act as a proof of $f \not\in \mathcal{C}$.
Further, for $D$ to be a non-trivial proof, its complexity needs to be significantly better than just enumerating over all circuits of size less than $s$.
It turns out that most known lower-bounds against algebraic models yield such ``distinguishers'' $D$ that are themselves computable efficiently, and hence they form the basis for \emph{algebraic natural proofs}.

More formally, a $\mathcal{D}$-natural proof for $\mathcal{C}$ is a polynomial $P \in \mathcal{D}$ that vanishes on the coefficient vectors of all polynomial in $\mathcal{C}$ (see \autoref{subsec:univ-circuits-nat-proofs} for details).
Motivated from algebraic geometry, sometimes such polynomials are also called \emph{equations for $\mathcal{C}$}.
Stated in these terms, almost every known algebraic circuit lower bound against a class $\mathcal{C}$ gives a $\VP$-natural proof for $\mathcal{C}$.
Here $\VP$ is the class of polynomials\footnote{Technically, the classes $\VP$, $\VNP$ contain polynomial families. We ignore this distinction for now for simplicity.} whose degree and circuit size grows polynomially with their arity.
Thus, asking whether natural techniques could prove a super-polynomial algebraic circuit lower bound, translates to asking: does $\VP$ have any $\VP$-natural proofs?

In this context, a work of Kumar, Ramya, Saptharishi and Tengse~\cite{KRST22} shows that if the Permanent is exponentially hard\footnote{Here exponential hardness means $2^{n^{\epsilon}}$-hardness for some constant $\epsilon > 0$.}, then there are no $\VP$-natural proofs for $\VNP$.
In particular, as the Permanent is widely believed to be exponentially hard, if one believes that $\VP$ has $\VP$-natural proofs, then they should expect $\VP \neq \VNP$ to be \emph{provable using natural techniques}. 
A natural question is therefore, what are the consequences of $\VP$ \emph{not admitting} any $\VP$-natural proofs?

\subsection*{Succinct Hitting Sets}

Suppose that the polynomials in $\mathcal{C}$ (their coefficient vectors, rather) form a \emph{hitting set} for the class $\VP$: low-degree polynomials with efficient circuits.
Just by definition, this means that no polynomial in $\VP$ can vanish over the entire class $\mathcal{C}$.
Stated simply, this would mean that the above \emph{natural} lower bound strategy is ineffective for the class $\mathcal{C}$.

Specifically, if $\mathcal{C}$ were to be the class of $n$-variate, multilinear polynomials with circuits of size $\poly(n)$, then the coefficient vectors have length $N = 2^n$.
If $\mathcal{C}$ forms a hitting set for $\VP(N)$, then it means that $\VP(N)$ has a hitting set that has an extremely succinct description: as the set of coefficient vectors of circuits of size about $\poly(n) = \poly(\log N)$.
Such a hitting set is therefore said to be a \emph{$C$-succinct hitting set} for $\VP$.
As argued above, the existence of $\VP$-succinct hitting sets for $\VP$ \emph{is equivalent to} the non-existence of $\VP$-natural proofs for $\VP$.
This is the core observation in the works of Forbes, Shpilka and Volk~\cite{FSV18}, and Grochow, Kumar, Saks and Saraf~\cite{GKSS17} (see \autoref{subsec:univ-circuits-nat-proofs} for formal statements).

Therefore, $\VP$ does not have $\VP$-natural proofs \emph{if and only if} algebraic circuits of size and degree $\poly(N)$ have hitting sets that can be described by circuits of size about $\poly(\log{N})$.
It is not hard to see that non-trivial hitting sets for a class almost immediately yield polynomials that hard for that class (see, for example, \cite{HS80}).
So what sort of hardness results do we get from these $\polylog(N)$-succinct hitting sets?
We show that existence of such hitting sets would imply that either $\VP \neq \VNP$ or that $\NC^1$ is separate from $\TC^0$ in a very strong sense (\autoref{thm:LBs-from-no-proofs}).

\paragraph*{Cryptographic Hitting Sets.}
On our way to proving this statement, we encounter an object that is weaker than ($\polylog(N)$) succinct hitting sets, but which we strongly believe is equally interesting on its own.
We call these objects \emph{cryptographic hitting sets}.

Informally, we say that a hitting set $H$ for a class $C$ is \emph{cryptographic}, if $H$ can be generated or described by a circuit in a class $C'$, where $C'$ is seemingly weaker than $C$.
For example, if $C$ is the class of algebraic circuits of size $s$, then the class of algebraic formulas of size $\poly(s)$, or even the class of circuits of size $t = s^{o(1)}$ would be valid choices for $C'$.
We define this concept formally as \autoref{def:crypto-hsg}.
The term \emph{cryptographic} here is chosen because this setting of parameters --- where a weaker class is required to fool a stronger class --- is prevalent in cryptography.

We then show an exciting consequence of cryptographic hitting set generators, that have a super-polynomial ``stretch''.
In particular, we prove that for the correct setting of parameters, if $\VP$-cryptographic HSGs exist for $\VP$ then either $\VP \neq \VNP$ or $\NC^1 \not \subseteq \TC^0$ (\autoref{thm:LBs-from-crypto-hsg}).

Note that we do not even know of a well-founded hardness assumption which implies the existence of cryptographic hitting sets for algebraic circuits.
The only example is a construction due to Kayal~\cite{K09}, whose setting of parameters is rather unwieldy for algebraic circuit complexity.
The work gives a polynomial map $\mathcal{M} : \F^N \rightarrow \F^{N+1}$ that can be described by constant depth formulas, such that any polynomial $P$ that vanishes on it: $P \circ \mathcal{M} \equiv 0$, has degree $2^{\Omega(N)}$.
The map is therefore a \emph{hitting set generator} for $N$-variate polynomials of degree $\poly(N)$.
This setting of parameters is unwieldy for two reasons: (1) the ``stretch'' of the generator is linear (can be made $N \rightarrow N+N^{1 + \epsilon}$ for any $\epsilon > 0$), and (2) we would ideally like a map that is hard to annihilate, \emph{despite} the existence of low-degree annihilators; this comes from the motivation to study ``syntactic'' algebraic computation, as outlined earlier.

\subsection{Our contributions}\label{subsec:our-results}

The main contribution of this paper are some interesting connections between lower bounds, identity testing and algebraic natural proofs.

\subsubsection*{Lower Bounds from Succinct Hitting Set Generators}

Before stating our theorem, let us recall that a family of polynomial maps $\setdef{H_n}{\C^n \to \C^{N(n)}}$ is a hitting set generator for a class of $N$-variate polynomials of degree-$D(N)$, if any family $\setdef{A_N}{A_N \in \F[x_1, \ldots, x_{N(n)}]}$ that annihilates $\set{H_n}$ is outside the class.
 
Succinct hitting set generators \cite{FSV18, GKSS17} are then defined as follows .
\begin{definition}[$\calc$-Succinct HSG for $\cald$]\label{defn:succ-hsg}
  For functions $d, D : \N \to \N$, let 
  \begin{itemize}\itemsep0pt
    \item $\calc_d$ be a class of $n$-variate polynomial families of degree $d(n)$, and
    \item $\cald_D$ be a class of $N$-variate polynomial families of degree $D(N)$ with $N = {n + d(n) \choose n}$.
  \end{itemize}
  We say that $\setdef{H_n}{\C^n \to \C^{N}}$ is a family of $\calc_d$-succinct hitting set generators for $\cald_D$, if the following are true.
  \begin{enumerate}\itemsep0pt
    \item $\set{H_n} \in \calc_d$.
    \item For infinitely many $n \in \N$, $H_n$ is a hitting set generator for $\cald_D(N)$. \qedhere
  \end{enumerate}
\end{definition}

Our main result shows that the existence of $\VP$-succinct hitting sets for $\VP$ (or equivalently, the existence of a barrier in proving $\VP \neq \VNP$ via \emph{natural techniques}) imply strong lower bounds.

\begin{restatable}[Hardness from Succinct Hitting Sets]{theorem}{SuccHSGtoLB}~\label{thm:LBs-from-no-proofs}
  Assuming the Generalized Riemann Hypothesis, if $\VP$-succinct hitting set generators exist for $\VP$, then at least one of the following must be true.
  \begin{enumerate}\itemsep0pt
    \item $\VP \neq \VNP$.
    \item For any $\ell(m) = o(1)$, there is a family of functions $\set{h_m}$ in uniform $\NC^1$ such that any uniform constant-depth threshold circuit computing it must have size larger than $\exp(\exp(\log^{\ell(m)}m))$, where the uniformity is $\mathsf{DLOGTIME}$ (in terms of the respective sizes).
  \end{enumerate}
  Further, if $\VP$ does not admit $\VNP$-natural proofs then $\P \neq\SPACE(\log ^{\log^*(n)}(n))$.
\end{restatable}

\begin{remark}
  A couple of remarks about \autoref{thm:LBs-from-no-proofs} are in order.
  \begin{enumerate}
    \item Comparing this result with that of Kabanets and Impagliazzo~\cite{KI04}, the (second) conclusion here is seemingly much stronger than $\mathsf{NEXP} \not\subseteq \byPoly{\P}$. This is to be expected, given that the hypothesis here demands a much more explicit derandomization of identity testing.
    \item Note that any function on $n$ variables in uniform $\NC^1$ is trivially computable by uniform constant-depth threshold circuits of size $\exp(\poly(n)) = \exp(\exp(O(\log n)))$; the statement in \autoref{thm:LBs-from-no-proofs} says that only a polynomial improvement in the second exponent is possible over this trivial bound.
    For comparison, $\exp(\exp(\log^{\Omega(1)}n))$ grows faster than a ``half-exponential'' function\footnote{The half-exponential function $h(n)$ is defined so that $h(h(n)) = 2^{n}$. For example, $ 2^{n^{\epsilon}} \gg h(n) \gg \operatorname{quasipoly}(n)$.}. 
    \qedhere
  \end{enumerate}
\end{remark}

\autoref{thm:LBs-from-no-proofs} essentially follows from the following upper bound on equations for evaluation vectors (similar to truth tables) of $\VP(n)$.

\begin{restatable}[Equations for \textsf{VP}]{theorem}{VPSPACEub}~\label{thm:vpspace-ub}
  For an arbitrary $d(n) \in \poly(n)$, let $N(n) = \binom{n + d(n)}{n}$.
  Then, for $t(n):= n^{\log^{\ast}n}$, there is a family of $N(n)$-variate, multilinear polynomials $\set{P_N}$ that depends only on the first $t(n)$ variables, satisfying the following.  
  \begin{itemize}\itemsep0em
    \item The family $\set{P_N}$ is a family of equations for \emph{the evaluation vectors} of $\VP_{d(n)}$.
    \item The coefficient functions of $\set{P_N}$ are computable in (uniform) space $t(n) = n^{\log^{\ast}n}$.
  \end{itemize}
\end{restatable}

\begin{remark}
  We make a few remarks about \autoref{thm:vpspace-ub}.
  \begin{enumerate}
    \item A $\VP(N)$\footnote{Here $\calc(n)$ and $\cald(N)$ mean that the complexities are polynomial in the parameters $n$ and $N \approx 2^n$, respectively.} upper bound on the equations of $\VP(n)$ would rule out a barrier in proving that $\VP \neq \VNP$ via natural proof strategies. 
    The bound we show is incomparable to $\VP(N)$.
    \item The function $\log^{\ast}(n)$ can be replaced by any growing function in $n$.
    \item Even though `algebraic natural proofs' are defined in terms of coefficient vectors instead of evaluation vectors, these notions are more or less equivalent.
    This is formalized as follows. 
    \begin{proposition}[Informal version of \autoref{prop:evaluation-and-coefficient-vectors}]~\label{prop:coeff-and-eval-eqns-informal}
      For any $d(n)$ and $N = \binom{n+d}{n}$, equations for \emph{coefficient vectors} of $\VP_{d}$ are computable in $\VP(N)$ (or $\VNP(N)$), if and only if, equations for \emph{evaluation vectors} of $\VP_d$ are computable in $\VP(N)$ (or $\VNP(N)$). \qedhere
    \end{proposition} 
  \end{enumerate}
\end{remark}

\subsubsection*{Defining Cryptographic Hitting Set Generators}
As mentioned earlier, our proof for \autoref{thm:LBs-from-no-proofs} goes via studying objects that are seemingly much weaker than succinct hitting sets. 
We now define them formally. 

Firstly, we define \emph{explicit polynomial maps} as follows.
\begin{definition}\label{defn:explicit-poly-maps}
  For a family $\setdef{H_n}{\C^n \to \C^{N}}$ of maps and a class $\calc$, we say that a circuit \emph{$C(\vecz,\vecy)$ encodes $H$} if there are assignments $\veca_1, \veca_2, \ldots, \veca_N$ to $\vecy$, such that for each $i \in [N]$, $C(\vecx,\veca_i) = h_i(\vecx)$.
  
  Analogously, we say that \emph{$\set{H_n}$ is $\calc$-explicit} if there is a family $\set{C_n} \in \calc$ such that, for every $n$, $C_n$ encodes $H_n$.
\end{definition}

\paragraph{Cryptographic HSGs} 
We are now ready to define cryptographic hitting set generators.

Recall that a polynomial map $H(x_1, \ldots, x_n) = (h_1(\vecx),\ldots,h_N(\vecx))$ is said to have degree $d$ if $h_i(\vecx)$ has degree at most $d$ for every $i \in [N]$. 

\begin{definition}[$\calc$-Cryptographic HSG for $\cald$]\label{def:crypto-hsg}
  For functions $d, D : \N \to \N$, let 
  \begin{itemize}
    \item $\calc_d$ be a class of $n$-variate polynomial families of degree $d(n)$, and
    \item $\cald_D$ be a class of $N$-variate polynomial families of degree $D(N)$.
  \end{itemize}
  Further, let $\setdef{H_n}{\C^n \to \C^{N(n)}}$ be a family of polynomial maps with $N(n) > n$.
  We say that $\set{H_n}$ is a family of $\calc_d$-cryptographic hitting set generators for $\cald_D$, if the following are true.
  \begin{enumerate}\itemsep0pt
    \item $\calc_d \subsetneq \cald_D$.
    \item $\set{H_n}$ is $\calc_d$-explicit.
    \item For infinitely many $n \in \N$, $H_n$ is a hitting set generator for $\cald_D(N)$. \qedhere
  \end{enumerate}
\end{definition}

\subsubsection*{Lower Bounds from Cryptographic HSGs}
We focus on $\VP$-explicit families of polynomial maps $\set{H_n}$, where the degree and size of encoding circuits increases polynomially in the number of parameters $n$, and show that hitting set generators for algebraic circuits that have a super-polynomial \emph{stretch} would have similar (but slightly weaker) consequences as the ones in \autoref{thm:LBs-from-no-proofs}.

\begin{restatable}[Lower Bounds from Cryptographic HSGs]{theorem}{CryptoHSGtoLB}\label{thm:LBs-from-crypto-hsg}
  Let $\setdef{H_n}{\C^n \to \C^{N}}$ be a family of polynomial maps of degree $d = n^{8}$ with $N \geq 2n$, and let $D(N) = N^{10}$.
  
  Assuming the Generalized Riemann Hypothesis, if the family $\set{H_n}$ is a $\VP_d$-cryptographic hitting set generator for $\VP_D$, at least one of the following must be true.
  \begin{enumerate}
    \item $\VP \neq \VNP$.
    \item Uniform $\NC^1 \not\subseteq$ uniform $\TC^0$, where the uniformity is $\mathsf{DLOGTIME}$.
  \end{enumerate}
  Further, if the family $\set{H_n}$ is a hitting set generator for $\VNP_d$, then $\P \neq \PSPACE$.
\end{restatable}

\begin{remark}\label{rmk:cryptographic-vp-vp}
  In \autoref{thm:LBs-from-crypto-hsg}, by ``$\set{H_n}$ is a $\VP_d$-cryptographic hitting set generator for $\VP_D$'' we mean that there is a fixed constant $e$ such that $\set{H_n}$ can be encoded using a size $n^e$ circuit, but $\set{H_n}$ hits circuits of size $N^a$ \emph{for all} constants $a$. This is what makes $\set{H_n}$ a cryptographic HSG.
\end{remark}

\subsubsection*{Annihilators of explicit polynomials maps}

All of our results follow from an upper bound that we are able to show on annihilators of explicit polynomial maps. 
The formal statement requires the notion of special evaluation gates, called \emph{projection gates} (see \autoref{defn:projection-gates}), and is stated as \autoref{thm:explicit-map-ann-ub}.

Projection gates characterize a class called $\VPSPACE$ in the same way as (usual) algebraic circuits characterize $\VP$.
A detailed discussion on this class can be found in \autoref{subsec:defining-VPSPACE}.
But over any fixed finite field, in particular, it coincides with the class of polynomials whose bits of coefficients can be computed in non-uniform $\PSPACE$.
In particular, a slightly weaker version of \autoref{thm:vpspace-ub} can be restated using this vocabulary as follows.

\begin{restatable}[Upper bound for annihilators of $\VP$]{theorem}{VPubAnn}\label{thm:vp-ann-ub}
  Let $d(m)$ be an arbitrary polynomial function of $m$. 
  For any $\VP_d$-explicit family of maps $\set{\calg_m}$, where each $\calg_m$ has $n$ outputs with $n \geq 2m$, there is a family $\set{A_m}$ in $\VPSPACE_b$ of degree $O(m^2d)$ such that $A_m$ annihilates $\calg_m$ for all large enough $m$.
\end{restatable}

In fact, the upper bound in this statement holds even when the encoding circuit uses projection gates (see \autoref{thm:vpspace-ann-ub}).
That is, $\VPSPACE$ even annihilates (low-degree) $\VPSPACE$-explicit maps\footnote{This can be phrased as `$\VPSPACE$ is closed under taking annihilators', in the same way as `$\VP$ is closed under taking homogenous components' and `$\VNP$ is closed under taking arbitrary coefficients'.}.
Using this, the hypothesis of \autoref{thm:LBs-from-crypto-hsg} can also be weakened to a family of $\VNP$ or even $\VPSPACE_b$-explicit HSGs (see \autoref{thm:vpspace-hsg-to-vp-neq-vpspace}).
It is therefore believable that \autoref{thm:vpspace-ub} can be improved; we expand on this in \autoref{subsec:open-questions}.
That said, as we have already seen, it is strong enough to have some very interesting consequences (\autoref{thm:LBs-from-no-proofs}, \autoref{thm:LBs-from-crypto-hsg}).

\subsection{Proof Overview}\label{subsec:proof-overview}
The main ideas behind our proofs are a combination of elementary techniques, and we will therefore try to give a nearly complete outline of our proofs.

\subsubsection*{Upper bound on the annihilators}

We begin with an overview of the proof of \autoref{thm:vp-ann-ub} and then discuss how \autoref{thm:vpspace-ub} follows from it.

Let $\mathcal{G} = (g_1(\vecz),\ldots,g_n(\vecz))$ be an arbitrary tuple of degree-$d$ polynomials in $m < n/2$ variables, that is $\VP_d$-explicit.
The explicitness of $\vecg$ will be used later in the proof.

Observe that an annihilator $A(x_1,\ldots,x_n) \in \C[\vecx]$ of \emph{individual} degree $(D-1)$ for this tuple is precisely a (non-trivial) $\C$-linear dependency between products of $g_1,\ldots,g_n$ of total degree at most $n(D-1)$, where each $g_i$ is multiplied at most $(D-1)$ times.
Now, any such product is itself a polynomial in $m$ variables of individual degree less than $d' := (nD \cdot d)$, and hence belongs to a vector space of dimension less than ${d'}^m$ over $\C$.
On the other hand, there are at least $D^n$ of these products, so if $D$ is large enough to ensure $D^n \geq {d'}^m = {(ndD)}^m$ then we are guaranteed some non-trivial linear dependency.
For $n \geq 2m$, this happens for a $D \leq nd$.

Now we turn to finding the coefficients of this dependency, which are going to be the coefficients of our annihilator.
For this, consider a matrix $M$ with rows labelled by monomials in $\vecz$ of individual degree at most $(d' - 1)$, and columns labelled by monomials in $\vecx$ of individual degree at most $(D-1)$.
Order both the rows and columns lexicographically.
For a monomial $\vecx^{\vece} = x_1^{e_1} \cdots x_n^{e_n}$, the $\vece^{th}$ column of $M$ is the coefficient vector of $\vecg^{\vece}$.
In this language, any column-dependency of $M$ is the coefficient vector of an annihilator of $(g_1,\ldots,g_n)$.

Since we are chasing an annihilator with as efficient a description as possible, it makes sense to pick the ``lexicographically first'' one.
That is, let $\vece$ be the ``smallest'' exponent such that $\vecg^{\vece}$ is dependent on the set of $\vecg^{\vece'}$s where all $\vece'$s are ``smaller than'' $\vece$.
Our annihilator will precisely be this unique dependency (up to scaling by $\C$).

Given a linear system of equations, $Ax = 0$, with a unique (up to scaling) solution, \emph{Cramer's rule} describes the solution in terms of determinants of submatrices of $A$.
We will fix the `last coefficient' to be $(-\det(A'))$, where $A'$ is the submatrix that has all but the last column.
This is done to ensure that all the coefficients are integers (and minors of $A$).
In fact, it turns out that stated this way, the annihilator itself can be written as a \emph{single} determinant just by inserting all the (symbolic) monomials $\vecx^{\vece}$ as the (new) last row!
We will call this `special Cramer's rule' for the remainder of this overview.

At this point, note that there is a minor issue.
For all we know, the ``first'' dependency that we have chosen gives us a submatrix (all columns up to $\vece$) that has more rows than columns, and hence we cannot apply Cramer's rule.
Further, it is not even clear if the basis of rows has a small description (smaller than just its characteristic vector).
We get around this by using a rank extractor, in particular the construction used by Forbes and Shpilka~\cite{FS12} (see \autoref{lem:rank-extractor}), because it fits snugly with the structure of the matrix $M$.
The arguments up to this point have been formalized in \autoref{lem:ann-as-det}.

Now that we are in the setting of `special Cramer's rule': we have an $(r+1) \times (r+1)$ matrix $\widetilde{M}$ whose determinant is the annihilator.
Since $r \leq {(md)}^{O(m)}$ this structure is not enough on its own to guarantee a non-trivial upper bound.
But we have even more structure in $\widetilde{M}$, owing to the explicitness of $\vecg$.
Specifically we have that, for all $i \leq r$, the ${(i,\vece_j)}^{th}$ entry of $\widetilde{M}$ is the evaluation of $\vecg^{\vece_j}$ at a point, $\beta^{(i-1)} \in \Z^{m}$ and this can be calculated efficiently by a constant-free algebraic circuit when $i$ is given in its binary representation.
Since computing a circuit for $\vecg^{\vece_j}$ given the binary encoding of $\vece_j$ is also doable efficiently, we say that the matrix $\widetilde{M}$ is efficiently encodable using algebraic circuits.
More formally, there is a size $\poly(\log{r}) = \poly(m,d,s)$ algebraic circuit which, when given bit-vectors corresponding to $i,j \in [r+1]$, outputs $\widetilde{M}[i,j]$.
Note that this output can sometimes be a monomial in $\vecx$.
The formal argument can be found in the proof of \autoref{lem:ann-as-succinct-det}.

Finally, we use the fact that the determinant of an explicit matrix can itself be computed in a ``memory-efficient'' way.
More formally, we use the elegant construction due to Mahajan and Vinay~\cite{MV97}, of an \emph{algebraic branching program} (see \autoref{defn:ABP}) for the determinant polynomial, along with repeated squaring using \emph{projection gates}  (see \autoref{defn:projection-gates}).
Here, the fact that projection gates let us evaluate a polynomial at constant cost proves to be crucial.
Quantitatively, given any matrix that is encoded by a circuit of size $s'$, we can compute the determinant of this matrix using a circuit with projection gates of size $\poly(s')$.
In fact, Malod~\cite{M11} uses this to give an alternative characterization of $\VPSPACE$.
This argument has been formalized in \autoref{prop:succinct-determinant-vpspace}.

\begin{figure}
  \begin{center}
    \resizebox{0.95\textwidth}{!}{
    \begin{tikzpicture}
      \draw (0,-1.5) rectangle (1,-1);
      \node at (0.5,-1.25) {$\vecx$};
      \draw [->] (0.1,-1) -- (0.1,1.5);
      \draw [->] (0.9,-1) -- (0.9,1.5);

      \draw (1.6,-1.5) rectangle (2.6,-1);
      \node at (2.1,-1.25) {$\vecu$};
      \draw [->] (1.7,-1) -- (1.7,0.5);
      \draw [->] (2.5,-1) -- (2.5,0.5);

      \draw (5.9,-1.5) rectangle (6.9,-1);
      \node at (6.4,-1.25) {$\vecv$};
      \draw [->] (6,-1) -- (6,0.5);
      \draw [->] (6.8,-1) -- (6.8,0.5);

      \node [rectangle,draw] at (4.25,0.15) {$z$};
      \draw [->] (4,0.1) to[out=180,in=270] (3.85,0.5);
      \draw [->] (4.45,0.1) to[out=0,in=270] (4.65,0.5);

      \node [rectangle,draw] at (4.25,-0.4) {$\text{ }\text{ }\vecw\text{ }\text{ }$};
      \draw [->] (3.75,-0.55) to[out=180,in=270] (2.95,0.5);
      \draw [->] (3.75,-0.35) to[out=180,in=270] (3.45,0.5);
      \draw [->] (4.75,-0.55) to[out=0,in=270] (5.55,0.5);
      \draw [->] (4.75,-0.35) to[out=0,in=270] (5.05,0.5);

      \draw (1.5,.5) rectangle (4,1);
      \node at (2.75,0.75) {$(1-z)\vecu + z\vecw$};
      \draw [->] (1.6,1) -- (1.6,1.5);
      \draw [->] (3.9,1) -- (3.9,1.5);

      \draw (4.5,.5) rectangle (7,1);
      \node at (5.75,0.75) {$(1-z)\vecw+z\vecv$};
      \draw [->] (4.6,1) -- (4.6,1.5);
      \draw [->] (6.9,1) -- (6.9,1.5);

      \draw (0,1.5) -- (7,1.5) -- (3.5,2.5) -- (0,1.5);
      \node at (3.5,2) {$C_i$};

      \draw [->] (3.5,2.5) -- (3.5, 2.95);
      \node [circle,draw] at (3.5,3.6) {$\mathsf{prod}_z$};
      \draw [->] (3.5,4.25) -- (3.5, 4.7);
      \node [circle,draw] at (3.5,5.35) {$\mathsf{sum}_{\vecw}$};

      \draw [->] (3.5,6) -- (3.5,7);
      \node at (4.75,6.75) {$D_{i+1}[\vecx,\vecu,\vecv]$};
        
      \draw [thick,dashed] (-0.25,-0.75) rectangle (7.25,6.25); 
      \node at (0.55,5.7) {$C_{i+1}$};

      \node at (3.5, -2) {\large{(1)}};

      \draw (8.75,-1) rectangle (9.75,-0.5);
      \node at (9.25,-0.75) {$\vecx$};
      \draw [->] (8.85,-0.5) -- (8.85,4.5);
      \draw [->] (9.65,-0.5) -- (9.65,4.5);

      \draw (10.25,-1) rectangle (11.25,-0.5);
      \node at (10.75,-0.75) {$\veci$};
      \draw [->] (10.35,-0.5) -- (10.35,4.5);
      \draw [->] (11.15,-0.5) -- (11.15,4.5);

      \draw (11.75,-1) rectangle (12.75,-0.5);
      \node at (12.25,-0.75) {$\vecj$};
      \draw [->] (11.85,-0.5) -- (11.85,4.5);
      \draw [->] (12.65,-0.5) -- (12.65,4.5);

      \draw (13.5,1) rectangle (15.5,2);
      \draw [->] (10.35,-0.5) to[out=45,in=270] (13.7,1);
      \draw [->] (11.15,-0.5) to[out=45,in=270] (14.7,1);
      \node [rectangle,draw,dashed] (alpha) at (15.1,1.5) {$\alpha$};
      \node at (14.525,2.3) {$\vecv_{\alpha,i}$};

      \draw (13.5,2.75) -- (15.5,2.75) -- (15.75,3.75) -- (13.25,3.75) -- (13.5,2.75);
      \draw [->] (13.65,2) -- (13.65,2.75);
      \draw [->] (15.35,2) -- (15.35,2.75);
      \node at (14.5,3.25) {$C_{\mathcal{G}}$};

      \draw (8.75,4.5) -- (15.75,4.5) -- (12.25,5.5) -- (8.75,4.5);
      \draw [->] (13.65,3.75) -- (13.65,4.5);
      \draw [->] (15.35,3.75) -- (15.35,4.5);
      \node at (14.55,4.1) {$C_{\mathcal{G}}(\vecv_{\alpha,i})$};

      \draw [->] (12.25,5.5) -- (12.25,6.5);
      \node at (13,6.25) {$M_{\mathcal{G}}[\veci,\vecj]$};
      
      \draw [thick,dashed] (8.5,.25) rectangle (16,5.75); 
      \node at (9,5.25) {$\widetilde{C}_{\mathcal{G}}$};

      \node at (12.25, -2) {\large{(3)}};
    \end{tikzpicture}}
  \end{center}

  \vspace{.25cm}
  \begin{center}
    \resizebox{0.75\textwidth}{!}{
    \begin{tikzpicture}
      \draw (3,2) -- (1.5,4.5) -- (0,2) -- (3,2);
      \node at (1.5,3) {$\widetilde{C}_{\mathcal{G}}$};
      \draw [->] (1.5,4.5) -- (1.5,5);

      \draw (3.5,-1.25) rectangle (4.5,-0.75);
      \node at (4,-1) {$\vecx$};
      \draw [->] (3.6,-0.75) -- (3.6,5);
      \draw [->] (4.4,-0.75) -- (4.4,5);

      \draw [thick,dashed] (5,-1.5) rectangle (6.5,-0.5);
      \draw (5.1,-1.1) rectangle (6.4,-0.6);
      \node at (5.3,-0.85) {$\ell_{\vecu}$};
      \draw (5.525,-1.1) -- (5.525,-0.6);
      \node at (5.75,-0.85) {$i_{\vecu}$};
      \draw (5.925,-1.1) -- (5.925,-0.6);
      \node at (6.2,-0.85) {$j_{\vecu}$};
      \node at (5.75,-1.325) {$\vecu$};
      \draw [->] (5.1,-0.5) -- (5.1,5);
      \draw [->] (6.4,-0.5) -- (6.4,5);

      \draw [thick,dashed] (7,-1.5) rectangle (8.5,-0.5);
      \draw (7.1,-1.1) rectangle (8.4,-0.6);
      \node at (7.3,-0.85) {$\ell_{\vecv}$};
      \draw (7.525,-1.1) -- (7.525,-0.6);
      \node at (7.75,-0.85) {$i_{\vecv}$};
      \draw (7.925,-1.1) -- (7.925,-0.6);
      \node at (8.2,-0.85) {$j_{\vecv}$};
      \node at (7.75,-1.325) {$\vecv$};
      \draw [->] (7.1,-0.5) -- (7.1,5);
      \draw [->] (8.4,-0.5) -- (8.4,5);

      \draw (12,2) -- (10.5,4.5) -- (9,2) -- (12,2);
      \node at (10.5,3) {$\widetilde{C}_{\mathcal{G}}$};
      \draw [->] (10.5,4.5) -- (10.5,5);

      \draw (0,5) -- (12,5) -- (6,6) -- (0,5);
      \draw [->] (6,6) -- (6,7);
      \node at (6.75,6.75) {$\mathcal{A}_{\mathcal{G}}[\vecu,\vecv]$};
      
      \draw [thick,dashed] (-.5,1) rectangle (12.5,6.25); 
      \node at (0.4,5.75) {$C'_{\mathcal{G}} = C_0$};

      \draw [->] (3.6,-0.75) to[out=150,in=270] (0.2,2);
      \draw [->] (4.4,-0.75) to[out=150,in=270] (1,2);
      \draw [->] (5.6,-0.6) to[out=140,in=270] (1.6,2);
      \draw [->] (5.8,-0.6) to[out=140,in=270] (1.9,2);
      \draw [->] (6.05,-0.6) to[out=130,in=270] (2.5,2);
      \draw [->] (6.25,-0.6) to[out=130,in=270] (2.8,2);

      \draw [->] (3.6,-0.75) to[out=70,in=270] (9.2,2);
      \draw [->] (4.4,-0.75) to[out=70,in=270] (10,2);
      \draw [->] (6.05,-0.6) to[out=50,in=270] (10.6,2);
      \draw [->] (6.25,-0.6) to[out=50,in=270] (10.9,2);
      \draw [->] (8.1,-0.6) to[out=60,in=270] (11.5,2);
      \draw [->] (8.3,-0.6) to[out=60,in=270] (11.8,2);

      \node at (6, -2.25) {\large{(2)}};
    \end{tikzpicture}}
  \end{center}
  \caption{The annihilator is computed by using (1) repeatedly to get $C_{\log N}$, if $N$ is the length of the ABP described by (2) (\autoref{claim:ABP-to-projection-circuit}); (2) describes the ABP computing $\det(\widetilde{M})$ using (3) (\autoref{claim:succinct-matrix-to-ABP}); (3) describes the matrix $\widetilde{M}$, such that $\det(\widetilde{M}) \circ \calg \equiv 0$, in terms of the generator $\calg$ (\autoref{lem:ann-as-succinct-det}).}
  \label{fig:ann-ckt}
\end{figure}
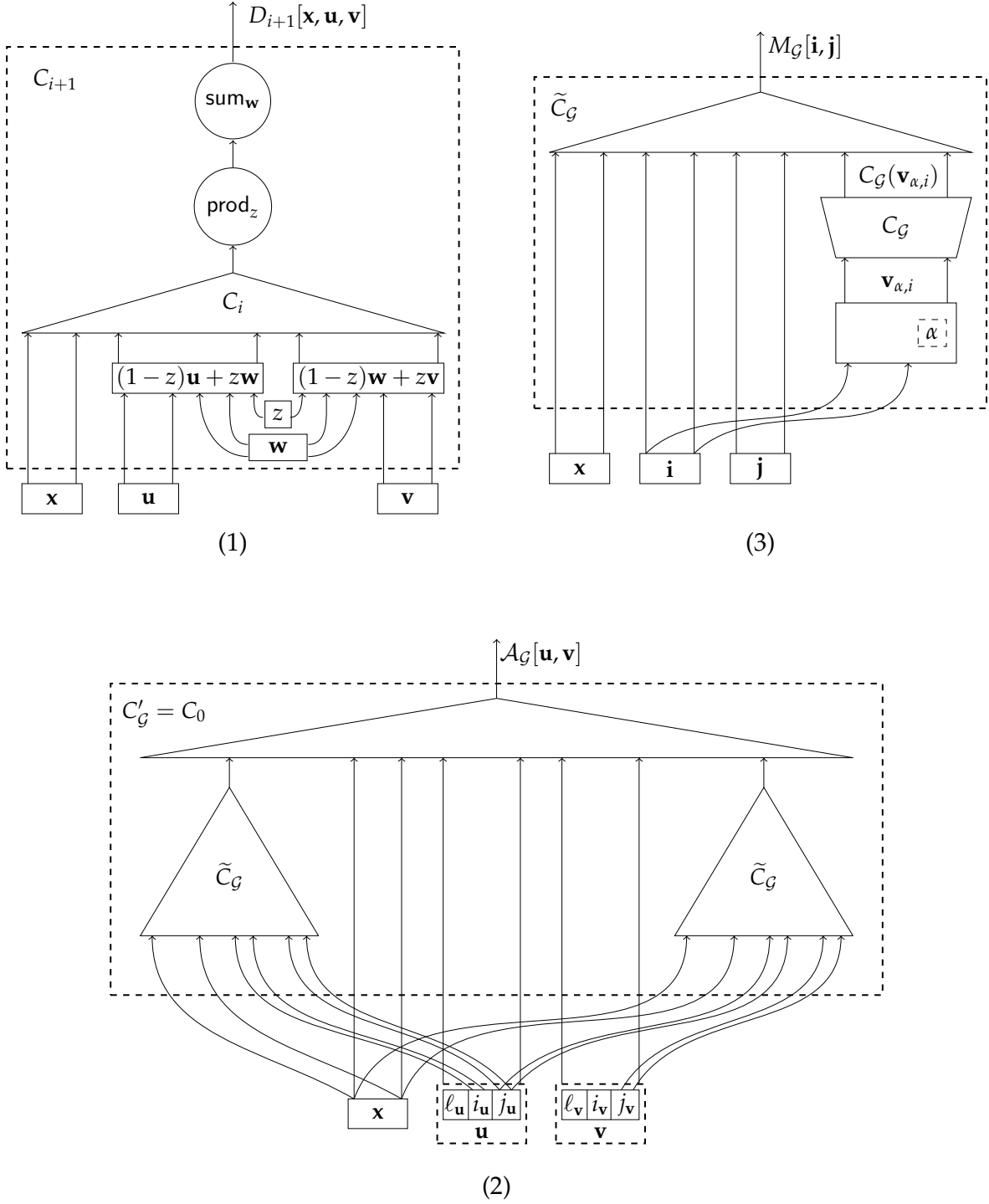

Putting all the pieces together (see \autoref{fig:ann-ckt}), we get a circuit with projection gates, of size $\poly(m,d,s)$ that computes the annihilator of the map $\mathcal{G} = (g_1,\ldots,g_n)$ as required.
A complete proof of \autoref{thm:vp-ann-ub} is given in \autoref{sec:vpspace-upper-bound}.

The observation that allows us to move from \autoref{thm:vp-ann-ub} to \autoref{thm:vpspace-ub} is that we can work with \emph{evaluation vectors} instead of \emph{coefficient vectors}, especially when dealing with $\VP$-natural or $\VNP$-natural proofs (due to \autoref{prop:coeff-and-eval-eqns-informal}).
This means that a family of equations can be constructed to vanish on the evaluation vector of the \emph{universal circuit} of a slightly super-polynomial size, say $n^{O(\log^{\ast}n)}$. 
Clearly, this evaluation vector is a $\size(n^{O(\log^{\ast}n)})$-explicit map.
Applying \autoref{thm:vp-ann-ub}, we get a family of equations that can be computed by circuits with projection gates, of size $n^{O(\log^{\ast}n)}$.
Finally applying the characterization of $\VPSPACE$ in terms of the computability of its coefficients (\autoref{defn:VPSPACE0-coefficient}) gives us the required statement. 
A formal proof of \autoref{thm:vpspace-ub} can be found in \autoref{subsec:vp-ann-ub}.

\begin{remark}
  Perhaps the first approach that one would take for proving \autoref{thm:vpspace-ub} would be to avoid using the notion of $\VPSPACE$.
  A sketch of this alternate proof is as follows.
  \begin{itemize}
    \item Given an $N \times N$ matrix with integer entries of bit-complexity $b$, there is an algorithm that uses space at most $\poly(\log{N},b)$, and simulates bit-access to its determinant \cite{B84, C76}.
    \item It can be shown that the definition of ``bit-complexity of the entries'' in the above result can be taken as ``space complexity of an algorithm that provides bit-access to the entries''.
    \item This when combined with the use of explicit rank-extractors \cite{GR05}, which makes the matrix $A$ have rank that is one less than full as well as have entries that have small ``bit-complexity'', essentially finishes the proof with an application of Cramer's Rule.
  \end{itemize}

  We believe that this alternate proof is relatively less clean, as it requires multiple switches between non-uniform algebraic computation and uniform boolean computation when one tries to formalize it fully.
  For instance, in the second step, the bound on the space complexity of the entries only holds when the map $\mathcal{G}$ itself is encoded by a \emph{constant-free algebraic circuit}, and showing that requires using the definition of $\VP$ as a \emph{projection} of the class\footnote{The class of polynomial families that are computable efficiently by constant-free algebraic circuits.} $\VP^0$.
  
  In comparison, our proof only uses non-uniform algebraic computation with the use of \emph{projection gates} and then uses the characterization of $\VPSPACE^0$ as (families of) polynomials whose coefficients are computable in non-uniform $\PSPACE$ (see \autoref{defn:VPSPACE0-coefficient}) only once at the the end. 
\end{remark}

\subsubsection*{Lower bounds from Hitting Set Generators}
\paragraph{Cryptographic HSGs}
An overview of how \autoref{thm:LBs-from-crypto-hsg} follows from \autoref{thm:vp-ann-ub} is the following.
Suppose, for contradiction, that $\VP = \VNP$ and uniform $\NC^1 \subseteq$ uniform $\TC^0$.
Further, let us assume that the Generalized Reimann Hypothesis is true.

It can be shown, using a padding argument (see, for example, \href{https://complexityzoo.net/Complexity_Zoo:C#ch}{Complexity Zoo}), that if uniform $\NC^1 \subseteq$ uniform $\TC^0$ then the counting hierarchy ($\CH$) is the same as $\PSPACE$.
This would then imply that $\byPoly{\CH} = \byPoly{\PSPACE}$.
On the other hand, since $\VP = \VNP$, $\byPoly{\CH} = \byPoly{\P}$.
Thus, we get that $\byPoly{\P} = \byPoly{\PSPACE}$, which would imply that $\VNP = \VPSPACE_b$ due to Valiant's Criterion and the definition of $\VPSPACE_b$.
Reusing the assumption that $\VP = \VNP$, we get that $\VP = \VPSPACE_b$.
For formal statements, see \autoref{prop:TC0_NC1-vs-CH_SPACE} and \autoref{prop:vp-neq-vpspace-counting-hierarchy}.

This means that any circuit with \emph{projection gates} can be efficiently simulated by a circuit without \emph{projection gates}.
In particular, this implies that the circuits which compute the family of annihilators of the hitting set generator $\set{\calh_m}$ can be converted to ``usual'' algebraic circuits of polynomially larger size.
Finally, since the degree function $d(n)$ is large enough for the annihilators ensured by \autoref{thm:vp-ann-ub}, we get that there is annihilator for $\set{\calh_m}$ in $\VP_d$.
This contradicts its `hitting property' in the hypothesis, thus completing the proof.

To prove the second statement, we assume for contradiction that $\P = \PSPACE$.
Thus, $\byPoly{\P} = \byPoly{\PSPACE}$ and therefore it must that $\VNP = \VPSPACE_b$.
An argument similar to the previous case then completes the proof.
\autoref{thm:LBs-from-crypto-hsg} is proved formally in \autoref{subsec:lb-from-crypto-hsg}.

\paragraph*{Succinct HSGs}
The only difference between Succinct HSGs and Cryptographic HSGs is the complexity of the circuit that encodes the generator.
Since the parameters are much better in the case of succinct HSGs --- barely super-polylogarithmic vs sub-polynomial in the number of outputs --- it allows us to extract much better lower bounds against constant-depth theshold circuits. 
A formal proof of \autoref{thm:LBs-from-no-proofs} is given in \autoref{subsec:lb-from-succ-hsg}.

\subsection{Open directions}\label{subsec:open-questions}

\begin{itemize}
  \item An intriguing takeaway from our results, is the case that it makes for studying `cryptographic HSGs' as in \autoref{thm:LBs-from-crypto-hsg} and the more general \autoref{thm:vpspace-hsg-to-vp-neq-vpspace}.
  Specifically, we ask whether the arguments of Heintz and Schnorr~\cite{HS80} can be derandomized within space that is sub-polynomial in the parameters $n$, $d$ and $s$.
  As this would satisfy the hypothesis of \autoref{thm:vpspace-hsg-to-vp-neq-vpspace}, achieving this even under a hardness assumption that is weaker than (or incomparable to) $\P \neq \PSPACE$ would be interesting.

  Along the same lines, under what hardness assumptions do `sub-exponentially strong cryptographic HSGs' corresponding to \autoref{thm:LBs-from-no-proofs} exist?
  There are some caveats\footnote{This discussion is beyond the scope of this paper. We encourage the reader to refer to the literature on algebraic natural proofs (e.g. \cite{AD08},\cite{FSV18}) for some details.} due to which boolean cryptographic primitives do not directly translate to HSGs.
  In particular, and specifically in the context of natural proofs, what would be an appropriate analogue for the work of \Hastad{}, Impagliazzo, Levin and Luby~\cite{HILL99}?

  \item Another interesting thread is to improve the upper bound in this paper on annihilators of explicit polynomial maps (\autoref{thm:vp-ann-ub}).
  Since we expect $\VPSPACE_b$ to be exponentially stronger than $\VP$ (and $\VNP$), any bound that is better than exponential, say e.g. $2^{o(s(n))}$ for size-$s(n)$-explicit maps, in terms of $\VP(2^n)$ (or $\VNP(2^n)$) would be interesting, even if it requires assuming some standard hypotheses.
  
  Annihilators from the `minimal dependency', as in \autoref{lem:ann-as-det}, are perhaps unlikely to imply such a bound without observing some structure of $\VP$.
  This is mainly because such an approach gives coefficients that are determinants of matrices of dimension $2^{\poly(s)}$ and one has to argue that all these coefficients can somehow be ``compressed'' into a smaller circuit.

  Note that \autoref{thm:vpspace-ann-ub} implies that the bound from \autoref{thm:vpspace-ub} extends to equations for $\VNP$.
  So, can we prove a better bound just for $\VP$?
  Certainly, deriving any property of $\VP$ that does not hold for $\VNP$ should probably assume hardness of $\VNP$ (e.g. \cite{KRST22}).
\end{itemize}

\subsection{Organisation of the paper}

The rest of our paper is organized as follows.

We first formalize some relevant concepts in \autoref{sec:prelim}. 
This section also includes some simple propositions (see \autoref{subsec:defining-VPSPACE}, \autoref{subsec:succinct-determinant-vpspace}) that will be important for the proof of our main theorems.
We then prove the $\VPSPACE$ upper bound on annihilators for $\VP$ (\autoref{thm:vp-ann-ub}) in \autoref{sec:vpspace-upper-bound} and our main theorems (\autoref{thm:LBs-from-crypto-hsg}, \autoref{thm:vpspace-ub}, \autoref{thm:LBs-from-no-proofs}) in \autoref{sec:consequences}.
For completeness, in \autoref{sec:vpspace-equiv-defns}, we include a somewhat detailed proof sketch of the equivalence of the two definitions of $\VPSPACE$ that we use.

\newpage

\section{Preliminaries}\label{sec:prelim}

\begin{itemize}
	\item We use $[n]$ to denote the set $\set{1,\ldots, n}$.

	\item We use boldface letters such as $\vecx, \vecy$ to denote tuples, typically of variables. When necessary, we adorn them with a subscript such as $\vecy_{[n]}$ to denote the length of the tuple. We also use $\vecx^{\vece}$ to denote the monomial $\prod x_i^{e_i}$.

	\item We use $\{f_n\}_{n \in \N}$ to denote families of polynomials. We drop the index set whenever it is clear from context. For a given polynomial $f$ we denote by $\deg(f)$ its degree. For a polynomial $f(\vecx,\vecy,\ldots)$ on multiple sets of variables, we use $\deg_{\vecx}(f) $, $\deg_{\vecy}(f) $, etc., to denote the degree in the variables from the respective sets.

	\item For a given $n$-variate polynomial $f(\vecx)$ of degree $d$, and a monomial $m$, we use $\coeff_{m}(f)$ to refer to the coefficient of $m$ in $f$. We further use $\cvector(f)$ to denote the vector\footnote{We do not explicitly mention the monomial ordering used for this vector representation, since all our statements work for any monomial ordering.} of coefficients of $f$.
	
  \item For a matrix $M$, we use $M[i,j]$ to refer to its $(i,j)^{th}$ entry. We commonly start our indices from $1$, but sometimes start them from $0$ when it helps the exposition. Whenever this is done, we make it clear.
\end{itemize}

\subsubsection*{Models of Algebraic Computation}
We now formally define the models of algebraic computation that we need for this work.

We first define algebraic circuits and their constant-free version.

\begin{definition}[Algebraic Circuits]\label{defn:algebraic-circuits}
  An \emph{algebraic circuit} is specified by a directed acyclic graph, with leaves (nodes with in-degree zero, called \emph{inputs}) labelled by field constants or variables, and internal nodes labelled by $+$ or $\times$.
  The nodes with out-degree zero are called the \emph{outputs} of the circuit. Computation proceeds in the natural way, where inductively each $+$ gate computes the sum of its children and each $\times$ gate computes the product of its children.

  The \emph{size} of the circuit is defined as the number of edges (or \emph{wires}) in the underlying graph.
\end{definition}

\begin{definition}[Constant-free Algebraic Circuits]~\label{defn:constant-free-circuits}
  An algebraic circuit is said to be ``constant-free'', if the only field constants that appear in it are $1$ and $-1$.
  In order that the circuit is able to compute rational numbers, we allow it to have division gates, where both the inputs are field constants.
\end{definition}

We also formally define algebraic branching programs.

\begin{definition}[Algebraic Branching Programs (ABPs)]\label{defn:ABP}
  An \emph{algebraic branching program} is specified by a layered graph where each edge is labelled by an affine linear form and the first and the last layer have one vertex each, called the ``source'' and the ``sink'' vertex respectively. 
  The polynomial computed by an ABP is equal to the sum of the weights of all paths from the start vertex to the end vertex in the ABP, where the weight of a path is equal to the product of the labels of all the edges on it. 
  
  The width of a layer in an ABP is the number of vertices in it and the width of an ABP is the width of the layer that has the maximum number of vertices in it.
  
  The size of an ABP is the number of edges in it. 
\end{definition}

\begin{remark}
  We will use ABPs to mean ABPs that have been defined as above, but with edge labels also being allowed to be monomials of degree that is at most the size of the ABP. 
\end{remark}

\subsubsection*{Algebraic Complexity Classes}
We now define the basic algebraic complexity classes.
\begin{definition}[$\VP$]\label{defn:VP}
	A family $\set{f_n}_n$ of polynomials is said to be in $\VP$, if there exists a constant $c \in \N$ such that for all large $n$, $f_n$ depends on at most $n^c$ variables, has degree at most $n^c$, and is computable by an algebraic circuit of size at most $n^c$.
\end{definition}
	
\begin{definition}[$\VNP$]\label{defn:VNP}
	A family $\set{f_n}_n$ of polynomials is said to be in $\VNP$, if there exists a constant $c \in \N$, and an $m$-variate family $\set{g_m} \in \VP$ with $m,\size(g_m) \leq n^c$, such that for all large enough $n$, $f_n $ satisfies the following.
	\[
		  f_n(\vecx) = \sum_{\veca \in \set{0,1}^{\abs{\vecy}}} g_{m}(\vecx,\vecy = \veca) \qedhere
	\]
\end{definition}

We will also be needing the following subclass of $\VP$, especially in the context of algebraic natural proofs.
\begin{definition}[$\VP_d$]\label{defn:d-slice-of-vp}
  For a function $d(n) \in \poly(n)$, we define $\VP_d$ to be the class of all degree-$d$ families that belong to $\VP$.
  Equivalently, a family $\set{f_n} \in \VP$ is said to be in $\VP_d$ if for all large enough $n$, $\deg(f_n) \leq d(n)$.

  We also define $\VNP_d$ to contain all the degree-$d$ families from $\VNP$.
\end{definition}

\paragraph*{Rank Extractor}
We will also need the concept of rank extractors.
A rank extractor is a family of maps which have rank preserving properties.
The following lemma shows the existence of such a family in the setting that we will need.
\begin{lemmawp}[\cite{FS12}]\label{lem:rank-extractor}
  Let $1 \leq r \leq n$ and let $M \in \C^{n \times r}$ be of rank $r$. 
  Define $A_\alpha \in \C^{r \times n}$ by $A_\alpha[i,j] = \alpha^{ij}$.
  Then there exists $\alpha \in [nr]$ such that $\rank(A_\alpha \cdot M) = r$.
\end{lemmawp}

\paragraph*{Polynomials for boolean operations}
Finally we note that there are computationally simple polynomials which capture certain boolean operations.
Since this is easy to check, we omit the proof. 

\begin{observationwp}\label{obs:boolean-operations}
  Given vectors $\veca,\vecb \in \set{0,1}^{\ell}$, the following polynomials (or vectors of polynomials) have constant-free circuits of size $O(\ell^2)$.
  \begin{itemize}\itemsep0pt
    \item $\mathrm{EQ}(\veca,\vecb)$: outputs $1$ if $\veca = \vecb$, $0$ otherwise.
    \item $\mathrm{GT}(\veca,\vecb)$: outputs $1$ if $\veca > \vecb$ when seen as binary encodings of positive integers, $0$ otherwise.
    \item $\mathrm{LT}(\veca,\vecb)$: outputs $1$ if $\veca < \vecb$ when seen as binary encodings of positive integers, $0$ otherwise.
    \item $\mathrm{INC}(\veca)$: outputs a vector $\vecb$ such that $\vecb = \veca + 1$, when seen as positive integers. \qedhere
  \end{itemize}
\end{observationwp}

\subsection{Universal Circuits and Natural Proofs}\label{subsec:univ-circuits-nat-proofs}

A circuit is said to be universal for circuits of size $s$ if every such circuit is a simple projection of it.

\begin{definition}[Universal Circuit~\cite{SY10}]\label{defn:universal-circuit-coeff-vector}
  A circuit $\mathcal{U}$ is called universal for $n$-input circuits of size $s$, that compute polynomials of degree $d$, if the following holds.
  \begin{quote}
    For any polynomial $f(x_1, \ldots, x_n)$ of degree $d$ that can be computed by a circuit of size $s$, there exists a circuit $\phi$ computing $f$ as well, such that the computation graph of $\mathcal{U}$ is the same as the graph of $\phi$. \qedhere
  \end{quote}
\end{definition}

The following lemma due to Raz \cite{R10a} shows the existence of such circuits.
For a proof sketch that yields the exact statement given below, please refer to \cite{CKRST20}.

\begin{lemmawp}[Existence of Universal Circuits~\cite{R10a}]\label{lem:size-of-universal-circuit}
	Let $\mathbb{F}$ be any field and $n,s\geq 1$ and $d \geq 0$.
  Then there exists an algebraic circuit $\mathcal{U}$ of size $\poly(n,d,s)$ computing a polynomial in $\mathbb{F}[x_1,\ldots,x_n,y_1,\ldots,y_r]$ with $r\leq \poly(n,d,s)$ such that:
	\begin{itemize}\itemsep0em
		\item $\deg_{\vecx}(\mathcal{U}(\vecx,\vecy)), \deg_{\vecy}(\mathcal{U}(\vecx,\vecy)) \leq \poly(d)$;
		\item for any $f\in\mathbb{F}[x_1,\ldots,x_n]$ with $\deg_{\vecx}(f)\leq d$ that is computable by an algebraic circuit of size $s$, there exists an $\veca\in \mathbb{F}^r$ such that $f(\vecx)= \mathcal{U}(\vecx,\veca)$. \qedhere
	\end{itemize}
\end{lemmawp}

\subsubsection*{Algebraic Natural Proofs}

We now define the concepts around natural proofs that we will need \cite{FSV18, GKSS17}.
\begin{definition}[Degree-$d$ families]\label{defn:deg-d-families}
  For any function $d : \N \rightarrow \N$, we denote by $\calp_d$ the class of all degree-$d$ polynomial families.
  That is, a family $\set{f_n}$ of $n$-variate polynomials belongs to $\calp_d$ if $\deg(f_n) \leq d(n)$ for all large enough $n$.
\end{definition}

\begin{definition}[$\cald$-Natural Proofs for $\calc$]
  Let $d(n), D(N)$ be polynomially growing functions, and let $\calc \in \calp_d$ and $\cald \in \calp_D$ be any classes of polynomial families.
  For $N(n) = \binom{n+d(n)}{n}$, we say that an $N$-variate family $\set{A_N} \in \cald$ is a \emph{$\cald$-natural proof for $\calc$} if for any family $\set{f_n} \in \calc$ we have that the polynomial $A_{N(n)}$ vanishes on the coefficient vector of $f_n$, for all large enough $n$.
\end{definition}

\begin{definition}[$\VP$, $\VNP$, and natural proofs]\label{defn:natural-proofs-vp-vnp}
  For a fixed $d(n) \in \poly(n)$, we say that \emph{$\VP$-natural proofs exist for $\VP_d$} if for some $D(N) \in \poly(N)$, there exists a family $\set{A_N}$ that is a $\VP_D$-natural proof for $\VP_d$.
  We say that \emph{$\VP$-natural proofs exist for $\VP$} if for every $d(n) \in \poly(n)$, there is some $\VP$-natural proof for $\VP_d$.

  Similarly, we say that there are \emph{$\VNP$-natural proofs for $\VP$}, if for each $d(n) \in \poly(n)$, there is some $\VNP$-natural proof.
\end{definition}

\subsubsection*{Succinct Hitting Sets}

For classes of polynomial families $\calc, \cald$, we have already defined $\calc$-succinct hitting set generators for $\cald$  (\autoref{defn:succ-hsg}). 
We now define $\VP$-succinct hitting set generators for $\VP$.
Note that they directly contradict $\VP$-natural proofs for $\VP$ \cite{FSV18, GKSS17}.
\begin{definition}[Succinct hitting sets for $\VP$]\label{defn:succinct-hsg-vp-vnp}
  For fixed $d(n) \in \poly(n)$, we say that \emph{$\VP_d$-succinct hitting sets exist for $\VP$} if for every $D(N), S(N) \in \poly(N)$, there is some family $\set{h_n}$ that is a $\VP_d$-succinct hitting set generator (as in \autoref{defn:succ-hsg}) for size-$s(N)$ families in $\VP_D$.

  We say that \emph{$\VP$-succinct hitting sets exist for $\VP$} if for some $d(n) \in \poly(n)$, there are $\VP_d$-succinct hitting sets for $\VP$.
  Finally, we say that there are \emph{$\VNP$-succinct hitting sets for $\VP$}, if for some $d(n) \in \poly(n)$, there are $\VNP_d$-succinct hitting sets for $\VP$.
\end{definition}

We refer the reader to the full version (on arxiv) of \cite{CKRST20} for detailed definitions of succinct HSGs and natural proofs.

\subsection{Algebraic analogues of bounded-space computation: \textsf{VPSPACE}}
\label{subsec:defining-VPSPACE}

The upper bound on the annihilator that we show is in terms of space complexity.
\begin{definition}[$\SPACE$]
  We use $\SPACE(s(n))$ to denote the class of languages that can be decided by deterministic Turing Machines in space $s(n)$ for all inputs of length $n$.
\end{definition}

\subsubsection*{Defining the class}
One way to define the algebraic analogue of $\PSPACE$ is using the space needed to compute coefficients of the polynomial, which was the route taken by Koiran and Perifel~\cite{KP09}.

\begin{definition}[Coefficient function]~\label{defn:coefficient-function}
  Suppose $f(x_1,\ldots,x_n)$ is a polynomial of \emph{individual} degree $d$, and each of its coefficient is an integer of absolute value $< 2^{\ell}$.
  
  Then, for $M = n \cdot \ceil{\log(d+1)} + (\ell+1)$, we define the coefficient function of $f$ to be the $M$-bit boolean function $\Phi_f$ such that $\Phi_f(e_1,\ldots,e_n,i)$ outputs the $i^{th}$ bit of the coefficient of the monomial $x_1^{e_1} \cdots x_n^{e_n}$, where the $0^{th}$ bit encodes the sign.
  
  A family $\set{f_n}$ of integer polynomials naturally defines a family $\set{\Phi_{n}}$ of coefficient functions.
\end{definition}

\begin{definition}[$\VPSPACE^0$ from coefficient functions~{\cite{KP09}}]~\label{defn:VPSPACE0-coefficient}
  A family $\set{P_N}$ of integer polynomials is said to be in $\VPSPACE^0$ if, for all large $N$, the polynomial $P_N$ has $\poly(N)$ variables, degree $2^{\poly(N)}$, each coefficient of $P_N$ has at most $2^{\poly(N)}$ bits, and if the family of coefficient functions of $\set{P_N}$ is computable in $\byPoly{\PSPACE}$.
\end{definition}

Poizat~\cite{P08b} on the other hand defined $\VPSPACE$ without going into boolean computation, using a new type of gate called projection gate, defined below.
He showed\footnote{Poizat's paper (written in French) includes a rough sketch of this proof. Malod's paper~\cite{M11} (in English) quotes his result but does not provide a proof.} that this definition is equivalent to that of Koiran and Perifel.

\begin{definition}[Projection gates]~\label{defn:projection-gates}
  A projection gate is labelled by a variable, say $w$, and a constant $b \in \set{0,1}$.
  We denote such a gate by $\proj_{w = b}$.
  The gate $\proj_{w = b}$ ``projects'' $w$ to $b$ in the input polynomial.
  That is, $\proj_{w = b} f(w,\vecx) = f(b,\vecx)$.
\end{definition}

\begin{definition}[Circuits with projections~{\cite{P08b}}]~\label{defn:VPSPACE0-projection}
  A family $\set{P_N}$ of integer polynomials is said to be in $\VPPROJ^0$ if for all large $N$, there is a constant-free algebraic circuit, say $C$, that additionally have access to projection gates such that $C$ has size $\poly(N)$ and $P_N$ is the polynomial computed by $C$.
\end{definition}

\begin{restatable}[{\cite{P08b,M11}}]{proposition}{VPSPACEequivalence}~\label{prop:equivalence-VPSPACE-definitions}
  $\VPSPACE^0 = \VPPROJ^0$.
\end{restatable}

We believe that this statement is not entirely obvious, so we give a fairly detailed proof sketch of this in \autoref{sec:vpspace-equiv-defns}.
Finally, for polynomials with arbitrary constants, $\VPSPACE$ is defined using $\VPSPACE^0$ as follows.

\begin{definition}[$\VPSPACE$]~\label{defn:VPSPACE}
  Let $\F$ be the field of rationals, reals or complexes.
  A family $\set{P_N}$ of polynomials over $\F$ is in $\VPSPACE$, if there exists an $M = \poly(N)$ and a family $\set{Q_M} \in \VPSPACE^0$, such that for all $N$, $P_N$ is obtained from $Q_M$ by setting some variables to constants from $\F$.
\end{definition}

For convenience, we also define two new types of gates: summation and production gates.
\begin{definition}[Summations and Productions]\label{defn:summations-productions}
  The summation and production operations are defined, using the operation of projection as defined in \autoref{defn:projection-gates}, as follows.
  \begin{itemize}\itemsep0pt
    \item $\summation_z f(z,\vecx) := \proj_{z=0} f(z,\vecx) + \proj_{z=1} f(z,\vecx)$
    \item $\production_z f(z,\vecx) := \proj_{z=0} f(z,\vecx) \times \proj_{z=1} f(z,\vecx)$ \qedhere
  \end{itemize}
\end{definition}
\noindent It is easy to see that, by definition, these gates can be simulated using projection gates and the usual sum, product gates.

Before moving on, it is important to note that there are (at least) two more equivalent characterizations of $\VPSPACE$ due to Malod~\cite{M11}, and Mahajan and Rao~\cite{MR13}.
We omit the details of these definitions as we do not directly use them.

\subsubsection*{Comparison with $\VP$}

Given that $\VPSPACE$ corresponds to a class as powerful as $\PSPACE$, it is worth checking that the known ``hard polynomial families'' are outside $\VPSPACE$ as well.
In this context, it makes sense to define a bounded degree analogue of $\VPSPACE$ written as $\VPSPACE_b$.
As Koiran and Perifel~\cite{KP09} showed, the degree bound is inconsequential when comparing $\VP$ and $\VPSPACE$.
\begin{propositionwp}[{Restatement of \cite[Lemma 4]{KP09}}]\label{prop:KP09-bounded-deg-vp-and-vpspace}
  Let $\VPSPACE_b$ be the polynomial families of degree $\poly(n)$ that belong to $\VPSPACE$, and similarly, let $\VP_{nb}$ be the class of polynomial families of unbounded degree that have algebraic circuits of size $\poly(n)$.

  Then, $\VP = \VPSPACE_b$ if and only if $\VP_{nb} = \VPSPACE$.
\end{propositionwp}

We can now summarize the relationship of $\VPSPACE_b$ with known hard polynomials as follows, essentially by using the arguments in the well-known book by {\Burgisser}~\cite{B00}.
\begin{itemize}\itemsep0em
  \item Since a \emph{random polynomial} does not have small circuits with projection gates with high probability, random polynomial families are outside $\VPSPACE_b$.
  \item Any construction that involves exponentially many ``independent irrational numbers'', e.g. $h_c(\vecx) = \sum_{\vece} \sqrt{p_{\vece}} \cdot \vecx^{\vece}$ for distinct primes $\set{p_{\vece}}$, also works against $\VPSPACE_b$.
  This is because the coefficients of polynomials in $\VPSPACE_b$ can always be written as integer polynomials that depend on at most polynomially many scalars from the underlying field.
  \item Finally, \emph{Strassen's multilinear polynomial}, defined\footnote{Here $\veci$ is the vector corresponding to the binary representation of $i$.} as $h_s(\vecx) = \sum_{0 \leq i < 2^n} 2^{2^i} \vecx^{\veci}$ has coefficients that are triply exponential in the number of variables $n$.
  This means that we will need doubly exponentially many bits to even index into the bits of the coefficients, which puts the (corresponding family of) coefficient functions outside $\PSPACE$.
\end{itemize}

\subsubsection*{Consequences of Separating $\VP$ and $\VPSPACE$}
Koiran and Perifel~\cite{KP09} showed that separating $\VP$ and $\VPSPACE_b$ would imply interesting lower bounds.

\begin{restatable}[{\cite[Proposition 3]{KP09}}]{propositionwp}{VPneqVPSPACEconsequenceKP}\label{prop:KP09-vp-neq-vpspace-full}
  If $\VP \neq \VPSPACE_b$, then either $\VP \neq \VNP$ or $\byPoly{\P} \neq \byPoly{\PSPACE}$.

  Assuming the Generalized Riemann Hypothesis (GRH), the converse is also true.
  That is, if $\VP \neq \VNP$ or $\byPoly{\P} \neq \byPoly{\PSPACE}$ then $\VP \neq \VPSPACE_b$. 
\end{restatable}

On the other hand, it can be shown via a padding argument that if $\TC^0 = \NC^1$ then the counting hierarchy ($\CH$) is the same as $\PSPACE$ (see, for example, \cite{CMTV98} or \href{https://complexityzoo.net/Complexity_Zoo:C#ch}{Complexity Zoo}) implying that $\byPoly{\CH} = \byPoly{\PSPACE}$ as well.
Further, the proof can be modified to show the following.

\begin{propositionwp}\label{prop:TC0_NC1-vs-CH_SPACE}
  There exists a constant $a$ for which the following holds.\\
  For any function $m(n) = \Omega(\log n)$, suppose that every $\operatorname{DTIME}(m(n))$-uniform $\NC^1$ circuit on $2^{a \cdot m(n)}$ inputs, can be simulated by a $\operatorname{DTIME}(n^c)$-uniform constant depth threshold circuit of size $2^{n^c}$ for some constant $c$.
  Then $\SPACE(m(n)) \subseteq \CH$, both defined for input length $n$.
\end{propositionwp}

At the same time, the proof of \autoref{prop:KP09-vp-neq-vpspace-full} can be generalized to extract the following statement when we additionally assume the GRH.
Here $\CH(t(n))$, $\VP(t(n))$, $\VPSPACE(t(n))$ refer to these classes defined with respect to the input length being $t(n)$.

\begin{proposition}\label{prop:vp-neq-vpspace-counting-hierarchy}
  Assuming the Generalized Riemann Hypothesis, for any function $m(n) = \Omega(\log n)$, if $\VPSPACE_b(m(n)) \not\subseteq \VP(n)$, then either $\VP \neq \VNP$ or $\SPACE(m(n)) \not\subseteq \CH(n)$.
\end{proposition}
\begin{proof}[Sketch.]
  The argument is easier for the contrapositive. We assume the GRH throughout.\\
  If $\VP = \VNP$, then from a work of \Burgisser{}~\cite{B09}, we know that $\CH$ collapses to $\byPoly{\P}$. Also, by Valiant's criterion~\cite{V79a}, we have that any polynomial family whose coefficient function (family) is in $\byPoly{\P}$ is in $\VNP$.
  
  Putting these together, we get that if $\VP = \VNP$, then any polynomial family whose coefficient function belongs to $\byPoly{\CH}$ must belong to $\VP$.
  However, if additionally $\SPACE(m(n)) \subseteq \CH(n)$, then all the polynomial families whose coefficient functions can be computed in (non-uniform) space $O(m(n))$ belong to $\VP$, which means that $\VPSPACE_b(m(n)) \subseteq \VP(n)$.
\end{proof}

\subsection{Explicit Objects}
Next, along the same lines as \emph{explicit} polynomial maps (\autoref{defn:explicit-poly-maps}), we define explicit matrices and explicit ABPs.
As before, we use the word \emph{explicit} to mean encodable by efficient circuits.

\begin{definition}[Explicit matrices]\label{defn:explicit-matrices}
  A circuit $C(\vecx, \veca, \vecb)$ is said to \emph{encode} a matrix $M \in \F^{r \times c}$ if $\abs{\veca} = \ceil{\log r}$, $\abs{\vecb} = \ceil{\log c}$ and for $\veci$, $\vecj$ being binary representations of $i$, $j$ respectively, $C(\vecx, \veci, \vecj) = M[i,j]$.

  Analogously, for a family of matrices $\setdef{M_{r,c}}{M_{r,c} \text{ has } r \text{ rows and } c \text{ columns}}$ and a class $\calc$, we say that \emph{$\set{M_{r,c}}$ is $\calc$-explicit} if there is a family $\set{C_{r,c}} \in \calc$ such that, for every $r,c$, $C_{r,c}$ encodes $M_{r,c}$.
\end{definition}

\begin{definition}[Explicit ABPs]\label{defn:explicit-ABPs}
  A circuit $C_A(\vecx, (\veca, \vecb), (\veca', \vecb'))$ is said to \emph{encode} an algebraic branching program $A$, of width $w$ and $d$ layers, if $\abs{\veca} = \abs{\veca'} = \ceil{\log d}$, $\abs{\vecb} = \abs{\vecb'} = \ceil{\log w}$ and  
  \begin{multline*}
    \forall \veci, \veci' \in \set{0, 1}^{\ceil{\log d}} \quad  \forall \vecj, \vecj' \in \set{0, 1}^{\ceil{\log w}},\\
    \quad C_A(\vecx, (\veci, \vecj), (\veci', \vecj')) \text{ is the label on the edge } ((i,j),(i',j')) \text{ in $A$}.
  \end{multline*}
  Here $\veci$, $\vecj$, $\veci'$, $\vecj'$ are the binary representations of $i$, $j$, $i'$, $j'$ respectively and $((i,j),(i',j'))$ denotes the edge between the $j$-th vertex in layer $i$ and the $j'$-th vertex in layer $i'$.

  Analogously, for a family of algebraic branching programs $\setdef{A_{w,d}}{A_{w,d} \text{ has width } w \text{ and } d \text{ layers}}$ and a class $\calc$, we say that \emph{$\set{A_{w,d}}$ is $\calc$-explicit} if there is a family $\set{C_{w,d}} \in \calc$ such that, for every $w,d$, $C_{w,d}$ encodes $M_{w,d}$.
\end{definition}

\subsection{Computing determinants of explicit matrices in \textsf{VPSPACE}}\label{subsec:succinct-determinant-vpspace}

We now show that given a matrix that is encoded by a small circuit, its determinant can be computed using a small circuit with projection gates.
A similar result already appears in a work by Malod~\cite{M11} and we provide a proof here for completeness.

\begin{proposition}[Computing determinant of explicit matrices]\label{prop:succinct-determinant-vpspace}
  Let $M \in \C^{N \times N}$ be a matrix that is encoded by a circuit $C(\vecx,\veca,\vecb)$ with $\abs{\veca}=\abs{\vecb}=\ceil{\log N}$.

  Then $\det(M)$ can be computed by a circuit $C'$ with projection gates, of size $O(\size(C) + \log^2 N)$.
  Moreover, if $C$ is constant-free, then so is $C'$.
\end{proposition}
\begin{proof}
  The proof has two parts.
  First we show that $\det(M)$ can be computed by an explicit Algebraic Branching Program (ABP) and then show that the polynomial computed by an explicit ABP can also be efficiently computed by circuits with projection gates.

  The first step is easy to deduce from the well-known elegant construction due to Mahajan and Vinay~\cite{MV97} and the second step is just a careful application of ``repeated squaring'' for matrices.
  We now prove these formally.

  \begin{claim}\label{claim:succinct-matrix-to-ABP}
    There is an explicit ABP, say $A(\vecx)$, encoded by a circuit of size $\poly(\size(C),\log N)$ that computes $\det(M)$.
  \end{claim}
  \begin{proof}
    We just describe the ABP computing $\det(M)$ here and also the circuit that encodes it.
    For a crisp proof of why this ABP computes the determinant, we direct the reader to Saptharishi's survey~\cite[Section 3.3.3]{S15}.

    The ABP has $N+1$ layers of vertices and, except for the first and the last layer, each layer has $O(N^2)$ vertices.
    Vertices of the ABP are labelled by $(\ell,(i,j))$ where $\ell \in [N+1]$ denotes the layer and $(i,j) \in [N] \times [N]$ indexes a particular vertex in that layer.
    The edges are given as below.
    \begin{itemize}\itemsep0em
      \item For each vertex $(\ell,(i,j))$ with $\ell \in [N-1]$, there is an edge to each $(\ell+1,(i,k))$ where $k > i$.
      The label of the edge is $M[j,k]$.
      \item For each vertex $(\ell,(i,j))$ with $\ell \in [N-1]$, there is an edge to each $(\ell+1,(k,k))$ where $k > i$.
      The label of the edge is $-M[j,i]$.
      \item All vertices $(N,(i,j))$ have an edge to the sink $(N+1,(1,1))$, with the label $-M[j,i]$.
    \end{itemize}

    We now describe the circuit $\widetilde{C}$ that encodes the above ABP, using the circuit $C$.
    Each vertex of the ABP is a vector of $3 \ceil{\log N}$ bits: $\ceil{\log N}$ each for $\ell$, $i$ and $j$ as described in the above construction.
    In the circuit, the two input vertices are $\vecu \equiv (\ell_u,(i_u,j_u))$ and $\vecv \equiv (\ell_v,(i_v,j_v))$\footnote{Here $\ell$,$i$,$j$ for both $\vecu$ and $\vecv$ are bit-vectors. We have used plain lowercase symbols since we think of them as numbers.}.
    Using the polynomials defined in \autoref{obs:boolean-operations}, we get the following expression for $\widetilde{C}$.
    \begin{align*}
      \text{Let } \mathrm{G}(\vecx,\vecu,\vecv) &= \mathrm{EQ}(\ell_v,\mathrm{INC}(\ell_u)) \cdot \mathrm{EQ}(i_u,i_v) \cdot \mathrm{GT}(j_v,i_u) \cdot C(\vecx,j_u,j_v)\\ 
      &+ \mathrm{EQ}(\ell_v,\mathrm{INC}(\ell_u)) \cdot \mathrm{GT}(i_v,i_u) \cdot \mathrm{EQ}(i_v,j_v) \cdot \inparen{-1 \cdot C(\vecx,i_u,j_u)}\\
      &+ \mathrm{EQ}(\ell_v,\mathrm{INC}(\ell_u)) \cdot \mathrm{EQ}(\ell_u,\vecn) \cdot \mathrm{EQ}(0 \ldots 01,i_v) \cdot \mathrm{EQ}(i_v,j_v) \cdot \inparen{-1 \cdot C(\vecx,i_u,j_u)},\\
      \text{and } \mathrm{valid}(\vecu) &= \mathrm{LT}(0\ldots 00,\ell_u) \cdot \mathrm{LT}(\ell_u,\mathrm{INC}(\mathrm{INC}(\vecn))) \cdot \mathrm{int}(i_u) \cdot \mathrm{int}(j_u),\\
      \text{where } \mathrm{int}(\veca) &= \mathrm{LT}(0\ldots 00,\veca) \cdot \mathrm{LT}(\veca,\mathrm{INC}(\vecn)).\\
      \text{Then } \widetilde{C}(\vecx, \vecu, \vecv) &= \mathrm{valid}(\vecu) \cdot \mathrm{valid}(\vecv) \cdot g(\vecx,\vecu,\vecv).
    \end{align*}
    Here each `product term' in the definition of $G$ corresponds to one type of edge in the description above, and $\vecn$, $0\ldots 00$ and $0\ldots 01$ are the bit-vectors corresponding to the numbers $N$, $0$ and $1$ respectively.
    We use the polynomial ``$\mathrm{valid}$'' to ensure that $\widetilde{C}$ outputs $0$ whenever it is given a pair of labels that is of a `non-edge'.

    The correctness of this expression is easy to check using the description of the ABP.
    Also, $\widetilde{C}$ has size $O(\size(C)+\log^2 N)$ and is a constant-free algebraic circuit (without projection gates) if $C$ is constant free.
  \end{proof}

  \begin{claim}\label{claim:ABP-to-projection-circuit}
    There is a circuit $C'$ with projection gates, of size $\poly(\size(\widetilde{C}),\log N)$ that computes the polynomial computed by the above ABP.
  \end{claim}
  \begin{proof}
    Let $A$ be the adjacency matrix of the graph that underlies the branching program.
    The polynomial computed by the ABP is just the $(s,t)$-th entry of the matrix $A^{N}$, where $s=(1,(1,1))$ and $t=(N+1,(1,1))$ are the tuples corresponding to the source and the sink.
    Since $\widetilde{C}$ encodes $A$, we only need to figure out how to encode $A^2$, which we can then use recursively to obtain access to the entries of $A^{2^k}$ for any $k$.
    Note that this becomes much easier when $N$ is a power of $2$.

    So we modify the ABP by adding layers $N+2,\ldots,N'+1$, to ensure that $N'$ is a power of $2$.
    Each of the new layers have a single vertex with the label $(\ell,(1,1))$ where $\ell$ is the layer.
    We also add an edge from $(\ell,1,1)$ to $(\ell+1,1,1)$, labelled with the scalar $1$, for every $N < \ell \leq N'$.
    This new ABP is explicit since we can add the term 
    \[
      \mathrm{EQ}(\ell_v,\mathrm{INC}(\ell_u)) \cdot \mathrm{GT}(\ell_u,\vecn) \cdot \mathrm{EQ}(0\ldots 01,i_u) \cdot \mathrm{EQ}(i_u,j_u) \cdot \mathrm{EQ}(0 \ldots 01,i_v) \cdot \mathrm{EQ}(i_v,j_v) \cdot 1
    \] 
    to the polynomial $G$ defined above, and also modify $\mathrm{valid}(\vecu)$ appropriately, to obtain a circuit that encodes the new ABP. 

    Now note that the $(s,t)$-th entry of $A^{N'}$ is the same as the $(s,t)$-th entry of $A^N$, and therefore is the polynomial we are looking for.
    Further, we can now easily compute the $(s,t)$-th entry of $A^{N'}$ using recursion, as described earlier.
    We now describe a circuit using projection gates that carries out this operation.
    
    Consider the following circuits defined using projections (\autoref{defn:projection-gates}) and summations (\autoref{defn:summations-productions}).
    \begin{align*}
      P_1(\vecx,\vecu,\vecw,\vecv,z) &:= \widetilde{C}\inparen{\vecx,((1-z)\vecu + z \vecw),((1-z)\vecw+z\vecv)}\\
      D_1(\vecx,\vecu,\vecv) &:= \summation_{w_1}\summation_{w_2}\cdots\summation_{w_L} \inparen{\proj_{z=0} P_1(\vecx,\vecu,\vecw,\vecv,z) \cdot \proj_{z=1} P_1(\vecx,\vecu,\vecw,\vecv,z)}
    \end{align*}
    Note that $D_1$ encodes the adjacency matrix $A^2$ when $\widetilde{C}$ encodes $A$ and $L = O(\log N)$ is the length of the vertex labels of $A$.

    Now for every $2 \leq i \leq k = \ceil{\log(N+1)}$, we define $P_{i+1}$ by using $D_i$ in place of $\widetilde{C}$, and $D_{i+1}$ by using $P_{i+1}$ in place of $P_1$, in the definitions of $P_1$ and $D_1$ above.

    As each of the $P_{i+1}$s and $D_{i+1}$s have circuits (with projection gates) of size $\poly(\log N)$ with \emph{exactly one} use of $D_{i}$ and $P_{i+1}$ respectively, all these increases are additive.
    So, finally, $C'(\vecx) = D_k(\vecx,s,t)$ has a circuit of size $O(\size(\widetilde{C}) + \log^2 N)$ and computes the polynomial we need.
    Further, $C'$ is constant-free if $\widetilde{C}$ is constant free.
  \end{proof}

  Combining both the above claims, we get that $\size(C') = O(\size(C) + \log^2 N)$, and also that $C'$ is constant-free whenever $C$ is constant-free.
\end{proof}

\section{\textsf{VPSPACE} Upper Bounds for Annihilators}\label{sec:vpspace-upper-bound}

In this section we prove \autoref{thm:vp-ann-ub}.
In fact we will prove the following, more general, statement.

\begin{restatable}[Annihilators of explicit maps]{theorem}{ExpMapAnnUB}\label{thm:explicit-map-ann-ub}
  Let $m$ be large enough, and let $\calg :\F^m \rightarrow \F^n$ be a polynomial map given by $\inparen{g_1(\vecz),\ldots,g_n(\vecz)}$ of degree $d$, with $n \geq 2m$.
  
  There exists a constant $c$ such that, if there is a circuit with \emph{projection gates} $C_{\calg}(\vecz,\vecy)$ of size $s$ that encodes $\calg$ as per \autoref{defn:explicit-poly-maps}, then there is a circuit \emph{with projection gates} $C'(\vecx)$ of size ${(m \cdot d \cdot s)}^c$ computing a nonzero polynomial $A(\vecx)$ of individual degree at most $3 \cdot m \cdot d$ that annihilates $\calg$ (that is, $A \circ \calg_m = A\inparen{g_1(\vecz),\ldots,g_n(\vecz)} \equiv 0$).
\end{restatable}

To prove this, we will first show that there is an \emph{explicit} matrix whose determinant is the annihilator, and then show that in this case the annihilator has a small circuit with projection gates.  

\subsection{Annihilator as determinant of a matrix}

\begin{lemma}[Annihilator as a determinant]\label{lem:ann-as-det}
  Let $\calg$ be a polynomial map as given in \autoref{thm:explicit-map-ann-ub}.
  Further, let $C_{\calg}(\vecz,\vecy)$ be a circuit that generates $\calg$.
  
  Then for $D = \ceil{{(nd)}^{m/(n-m)}} + 1$, there exists a $K \times K$ matrix $\widetilde{M}$ with $K \leq D^n$ such that $\det(\widetilde{M})$ is an annihilator of $\calg$ that has individual degree at most $D-1$.
  Moreover, $\widetilde{M}$ can be described as follows.

  There is some positive integer $\alpha < D^{2n}$ such that, for $i \in \set{0, \ldots, K-1}$ and $j \in [K]$,
  \[
    \widetilde{M}[i,\vece^{(j)}] = 
    \begin{cases}
      (\mathcal{G}(\vecv_{\alpha,i}))^{\vece^{(j)}} & \text{ if } i \leq K-2\\
      (-1)^{K-1} \cdot \vecx^{\vece^{(j)}} & \text{ if } i=K-1,
    \end{cases} 
  \]
  where $\vecv_{\alpha,i} = (\alpha^{i}, \alpha^{i \cdot ndD}, \ldots, \alpha^{i \cdot (ndD)^{m-1}})$.
\end{lemma}

\begin{proof}
  First we see how linear dependencies of specific polynomials in terms of $\calg$ give us annihilators of the map $\calg$.
  Suppose $A(\vecx)$ is an annihilator of $\calg$, and let $A(\vecx) = \sum_{\vece} f_{\vece} \vecx^{\vece}$ be its sparse representation.
  Then $0 = A(\calg) = \sum_{\vece} f_{\vece} \calg^{\vece}$.
  In other words, the coefficient vector of $A$ is a linear dependency in the set of polynomials of the form $\calg^{\vece}$, as we range over all the exponent vectors $\vece$.
  Clearly, the converse is also true: for any polynomial $A(\vecx) = \sum_{\vece} f_{\vece} \vecx^{\vece}$, if the coefficients $\set{f_{\vece}}$ represent a linear dependency in the polynomials $\set{\calg^{\vece}}$, then $F(\calg) = 0$.

  Now, for a parameter $D$ to be fixed later, consider the following matrix $M$.
  \begin{itemize}\itemsep0pt
    \item Columns of $M$ are indexed by monomials in $\vecx = \set{x_1,\ldots,x_n}$ of individual degree at most $D-1$, ordered lexicographically.
    \item Rows of $M$ are indexed by monomials in $\vecz = \set{z_1,\ldots,z_m}$ of individual degree at most $(ndD) - 1$, ordered lexicographically.
    \item For any valid $\vece \in \N^n$, the $\vece$th column of $M$ is the coefficient vector of the product $\calg^{\vece} = {g_1(\vecz)}^{e_1} \cdots {g_n(\vecz)}^{e_n}$.
    Note that this product is an $m$-variate polynomial of individual degree at most $\abs{\vece} \leq (n \cdot (D-1)) \cdot d < (ndD) - 1$.
  \end{itemize}

  From the previous discussion, we can see that any nonzero dependency in the columns of $M$ is the coefficient vector of some annihilator of $\calg$.
  Now observe that $M$ has fewer rows than columns whenever $D > {(nd)}^{m/(n-m)}$, since ${(ndD)}^m < D^n$ in this case.
  Therefore, there is a non-trivial dependency in its columns if we set $D = \ceil{{(nd)}^{m/(n-m)}} + 1$.

  In order to ensure a \emph{unique} dependency (up to scaling), we restrict $M$ to its first $K$ columns, where $K-1$ is the largest number for which the first $K-1$ columns are linearly independent.
  So now $M$ has the following properties.
  \begin{itemize}\itemsep0pt
    \item M has $K \leq D^n$ columns and each column is labelled by a monomial in $\set{x_1, \ldots, x_n}$ of individual degree $\leq D-1$.
    Let $\set{\vece^{(j)}}_{j\in[K]}$ be the labels of the columns.
    \item Each row is labelled by a monomial in $\set{z_1, \ldots, z_m}$ of individual degree $\leq (ndD) - 1$ and so, if the number of rows in $M$ is $R$, then 
    \[
      R \leq (ndD)^m < D^n
    \]
    \item The $(\vecz^\vece, \vece^{(j)})$-th entry of $M$ is the coefficient of ${\vecz^\vece}$ in $\mathcal{G}^{\vece^{(j)}}$.
    \item The first $K-1$ columns are linearly independent and the last column is a linear combination of the previous columns.
    \item $A(\vecx) = \vecx^{\vece^{(K)}} - \sum_{j=1}^{K-1} f_{\vece^{(j)}} \cdot \vecx^{\vece^{(j)}}$ is an annihilator of $\mathcal{G}$ ,if $M[\vece^{(K)}] = \sum_{j=1}^{K-1} f_{\vece^{(j)}} \cdot M[\vece^{(j)}]$.
  \end{itemize}  

  We now construct a matrix $M'$, with $(K-1)$ rows and $K$ columns, such that $M'$ has rank exactly $(K-1)$.
  In fact, the columns of $M'$ will share the same dependencies as the columns of $M$. 

  \begin{claim}
    Define for $\alpha \in \N$, the matrix $E_\alpha \in \C^{(K-1) \times R}$ such that $E_\alpha[i,j] = \alpha^{ij}$ for each $0 \leq i < K-1$ and $0 \leq j < R$.
    For some $\alpha \leq D^{2n}$, the matrix product $M' := E_\alpha \cdot M$ has rank exactly $K-1$.
    Further, $M'$ has the same dependencies in its columns as $M$.
  \end{claim}

  \begin{proof}
    By \autoref{lem:rank-extractor}, we immediately have that there exists $\alpha \leq R \cdot (K-1) < D^{2n}$ such that $\rank(E_\alpha \cdot M) = K-1$.
    Let $M' = E_\alpha \cdot M$.
    To show that $M'$ has the same dependencies in its columns as $M$, let us first assume that $M\vecu = 0$ for some $\vecu \in \C^K$.
    Clearly this implies that $M'\vecu = E_\alpha \cdot M \vecu = 0$.
    
    Conversely, let $\vecv \in \C^K$ be such that $M'\vecv = 0$ but $M \vecv \neq 0$.
    Then, clearly, $\vecv \neq \beta \vecu$ for any $\beta \in \C$.
    This shows that $\dim(\ker(M')) \geq 2$, contradicting the rank-nullity theorem since $\rank(M') = K-1$.
    Therefore, no such $\vecv$ can exist, proving that for any $\vecv \in \C^K$, $M'\vecv = 0 \implies M \vecv = 0$.
  \end{proof}

  For $M'$ defined as in the claim above, we clearly have that $M'[\vece^{(K)}] = \sum_{j=1}^{K-1} f_{\vece^{(j)}} \cdot M'[\vece^{(j)}]$.
  Also, note that the rows of $M'$ are labelled by $\set{0, \ldots, K-2}$ and the columns are labelled by $\setdef{\vece^{(j)}}{j\in[K]}$.
  We now define $\widetilde{M}$, a $K \times K$ matrix with rows labelled by $\set{0, \ldots, K-1}$ and columns labelled by $\setdef{\vece^{(j)}}{j\in[K]}$, as follows.
  \[
    \widetilde{M}[i,\vece^{(j)}] = \begin{cases}
      M'[i,\vece^{(j)}] & \text{ if } i \leq K-2\\
      (-1)^i \cdot \vecx^{\vece^{(j)}} & \text{ if } i=K-1
    \end{cases} 
  \]

  First, we show that the determinant of $\widetilde{M}$ is an annihilator of $\mathcal{G}$. 
  In particular, we show that $\det(\widetilde{M})$ is a scalar multiple of $A(\vecx)$.  
  We use $M'[\ast,\vece^{(j)}]$ to denote the $\vece^{(j)}$th column of the matrix $M'$, and $M'\setminus M'[*,\vece^{(j)}]$ to denote the submatrix of $M'$ obtained by deleting the $\vece^{(j)}$th column.
  
  \begin{claim}
    $\det(\widetilde{M}) = \det(M'\setminus M'[*,\vece^{(K)}]) \cdot A(\vecx)$.
  \end{claim}

  \begin{proof}
    Note that, by expanding with respect to the last row,
    \[
      \det(\widetilde{M}) = \sum_{j=1}^{K} (-1)^{j-1} \cdot \det(M'\setminus M'[*,\vece^{(j)}]) \cdot (-1)^{K-1} \cdot \vecx^{\vece^{(j)}}.
    \]
    This can be re-written as
    \[
      \frac{\det(\widetilde{M})}{\det(M'\setminus M'[*,\vece^{(K)}])} = \vecx^{\vece^{(K)}} + \sum_{j=1}^{K-1} (-1)^{K-j} \cdot \frac{\det(M'\setminus M'[*,\vece^{(j)}])}{\det(M'\setminus M'[*,\vece^{(K)}])} \cdot \vecx^{\vece^{(j)}}.
    \]
    Since $M'[\vece^{(K)}] = \sum_{j=1}^{K-1} f_{\vece^{(j)}} \cdot M'[\vece^{(j)}]$, it is now easy to check the following using Cramer's Rule. 
    \[
      \frac{\det(\widetilde{M})}{\det(M'\setminus M'[*,\vece^{(K)}])} = \vecx^{\vece^{(K)}} - \sum_{j=1}^{K-1} f_{\vece^{(j)}} \cdot \vecx^{\vece^{(j)}} = A(\vecx). \qedhere
    \]
  \end{proof}

  Since $\det(M'\setminus M'[*,\vece^{(K)}])$ is a fixed, nonzero scalar, $\det(\widetilde{M})$ is an annihilator of $\mathcal{G}$.
  Further, since every monomial in $\setdef{\vecx^{\vece^{(j)}}}{j\in[K]}$ has individual degree at most $D-1$, $\det(\widetilde{M})$ also has individual degree at most $D-1$.

  We now show that $\widetilde{M}$ can be described as claimed.
  To show that, we note that $M'$ has a very special structure.
  For any $\ell \in \set{0, \ldots, R-1}$, let $\vecv_\ell \in [ndD-1]^m$ be the unique vector such that 
  \[
    \ell = \sum_{b=1}^{m} \vecv_\ell(b) \cdot (ndD)^{b-1}.
  \]
  Then, for $i \in \set{0,\ldots,K-2}$ and $j \in [K]$,
  \begin{align*}
    M'[i,\vece^{(j)}] &= \sum_{\ell=0}^{R-1} \alpha^{i\ell} \cdot \coeff_{\vecz^{\vecv_\ell}}(\mathcal{G}^{\vece^{(j)}}) = \sum_{\ell=0}^{R-1} \coeff_{\vecz^{\vecv_\ell}}(\mathcal{G}^{(\vece^{j})}) \cdot \alpha^{i \cdot \inparen{\sum_{b=1}^{m} \vecv_\ell(b) \cdot (ndD)^{b-1}}}\\
    &= \sum_{\ell=0}^{R-1} \inparen{\coeff_{\vecz^{\vecv_\ell}}(\mathcal{G}^{\vece^{(j)}}) \prod_{b=1}^{m} \alpha^{(i \cdot (ndD)^{b-1}) \vecv_\ell(b)}} 
    = \sum_{\ell=0}^{R-1} \inparen{\coeff_{\vecz^{\vecv_\ell}}(\mathcal{G}^{\vece^{(j)}}) \cdot \vecv_{\alpha,i}^{\vecv_\ell}} 
  \end{align*}
  where $\vecv_{\alpha,i} = (\alpha^i, \alpha^{i \cdot ndD}, \ldots, \alpha^{i \cdot (ndD)^{m-1}})$.
  That is, 
  \[
    M'[i,\vece^{(j)}] = (\mathcal{G}(\vecv_{\alpha,i}))^{\vece^{(j)}}.
  \]
  We can therefore re-write $\widetilde{M}$ as claimed.
\end{proof}

\subsection{Annihilator as determinant of an explicit matrix}
We now show that the matrix described in the last section is \emph{explicit} in the sense of \autoref{defn:explicit-matrices}.
  
\begin{lemma}[Annihilator as determinant of an explicit matrix]\label{lem:ann-as-succinct-det}
  Let $\calg$ be a polynomial map as given in \autoref{thm:explicit-map-ann-ub}.
  Let $C_{\calg}(\vecz,\vecy)$ be a circuit that encodes $\calg$.
  Further, for $D = \ceil{{(nd)}^{m/(n-m)}} + 1$, $\alpha < D^{2n}$ and $K \leq D^n$ , let $\widetilde{M}$ be defined as follows.

  For $i \in \set{0, \ldots, K-1}$ and $j \in [K]$,
  \[
    \widetilde{M}[i,\vece^{(j)}] = 
    \begin{cases}
      (\mathcal{G}(\vecv_{\alpha,i}))^{\vece^{(j)}} & \text{ if } 0 \leq i \leq K-2\\
      (-1)^{K-1} \cdot \vecx^{\vece^{(j)}} & \text{ if } i=K-1.
    \end{cases} 
  \]
  where $\vecv_{\alpha,i} = (\alpha^{i}, \alpha^{i \cdot ndD}, \ldots, \alpha^{i \cdot (ndD)^{m-1}})$.

  Then, given access to a $C_{\calg}$ gate, there is a \emph{purely algebraic circuit}\footnote{Algebraic circuit without projection gates.} $\widetilde{C}_{\calg}$ that encodes $\widetilde{M}$.
  Further, $\widetilde{C}_{\calg}$ uses only a single $C_{\calg}$ gate and has overall size $O((nd)^3 + n \cdot \size(C_{\calg}))$.
\end{lemma}

\begin{proof}
  Let $\vecj$ be a bit-vector of length $n \cdot \delta$, for $ \delta := \ceil{\log D}$, which we interpret as the tuple $(\vecj^{(1)},\ldots,\vecj^{(n)})$ where $\vecj^{(\ell)}$ is the bit-vector encoding the $\ell$-th co-ordinate of $\vece^{(j)}$.
  Further, let the bit-vector $\veci$ encode the integer $i$.
  We want to construct a circuit $\widetilde{C}_{\calg}$ such that $\widetilde{C}_{\calg}(\vecx, \veci, \vecj) = \widetilde{M}[i,\vece^{(j)}]$.
  
  Firstly, it is easy to see that there is a constant-free multi-output circuit $C'$, of size $O(\log^2 K)$, such that $C'(\vecj) = (\vecj^{(1)},\ldots,\vecj^{(n)})$ as described above.
  We will use $C'_{k,b}(\vecj)$ to denote the output gate of $C'(\vecj)$ which outputs the $b$-th bit of $\vecj^{(k)}$.
  
  Let $\Delta = ndD$ and $\vecv_{\alpha} := (\alpha, \alpha^{\Delta}, \ldots, \alpha^{\Delta^{m-1}})$.
  We start by computing $L = \ceil{\log K}$ many distinct powers of the vector $\vecv_{\alpha}$, namely, $\vecv_{\alpha}$, $\vecv_{\alpha}^2$, $\vecv_{\alpha}^4$, $\ldots$, $\vecv_{\alpha}^{2^{L-1}}$.
  Here $\vecv_{\alpha}^r$ is used to denote the vector $(\alpha^r, \alpha^{r \cdot \Delta}, \ldots, \alpha^{r \cdot \Delta^{m-1}})$.
  Since $\alpha$ can be computed by a constant-free circuit of size $O(n \log(nd))$, each of these $L$ vectors can be computed by a constant-free circuit of size $O(L \cdot n \log(nd)) = O(\log K \cdot n \log(nd))$.
  Let these multi-output circuits be $C_0(\alpha), \ldots, C_{L-1}(\alpha)$ with $C_{\ell,k}(\alpha)$ denoting the $k$-th output gate in $C_\ell(\alpha)$ for $k \in \set{0, \ldots, m-1}$.
  That is, $C_{\ell,k}(\alpha) = \alpha^{r \cdot \Delta^k}$ for $k \in \set{0, \ldots, m-1}$.

  We now describe some intermediate circuits, and then finally $\widetilde{C}_{\calg}$.
  For $\veci = (i_0, \ldots, i_{L-1})$, 
  \begin{align*}
    \mathrm{pow}(\veci) &:= \inparen{\prod_{\ell = 0}^{L - 1} \inparen{i_{\ell} \cdot C_{\ell,0}(\alpha) + (1 - i_{\ell}) \cdot 1 }, \ldots, \prod_{\ell = 0}^{L - 1} \inparen{i_{\ell} \cdot C_{\ell,m-1}(\alpha) + (1 - i_{\ell}) \cdot 1 }}\\
    \mathrm{ROW}(\veci) &:= \mathrm{LT}(\veci,\veck) \cdot C_\calg(\mathrm{pow}(\veci)) + \mathrm{EQ}(\veci,\veck) \cdot \vecx\\
    \widetilde{C}_{\calg}(\veci,\vecj) &:= \prod_{a \in [n]} \inparen{ \prod_{b = 0}^{\delta - 1} \inparen{C'_{a,b}(\vecj) \cdot (\mathrm{ROW}(\veci)_a)^{2^b} + (1 - C'_{a,b}(\vecj)) \cdot 1} }
  \end{align*}

  Here $\mathrm{pow}(\veci)$ computes the vector $\vecv_{\alpha,\veci}$, and the polynomials $\mathrm{LT}(\cdot,\cdot)$ and $\mathrm{EQ}(\cdot,\cdot)$ are as defined in \autoref{obs:boolean-operations}.
  Further, $\veck$ is the binary encoding of the integer $K-1$ and $\mathrm{ROW}(\veci)_a$ denotes the $a$-th output gate of $\mathrm{ROW}(\veci)$.
  That is, $\mathrm{ROW}(\veci)_a = \mathrm{LT}(\veci,\veck) \cdot g_a(\mathrm{pow}(\veci)) + \mathrm{EQ}(\veci,\veck) \cdot x_a$ if $\mathcal{G} = (g_1, \ldots, g_n)$. 

  Note that $\widetilde{C}_{\calg}$ uses the sub-circuit $C_{\calg}$ exactly once, in $\mathrm{ROW}$, as claimed.
  Also, all the three expressions described above are constant-free, algebraic expressions.
  Finally, note that the size of $\widetilde{C}_{\calg}$ is $O(L \cdot n \log(nd) + L + L^2 + nd + n \cdot \size(C_{\calg}))$, which is $O(n^3 \cdot d^2 + n \cdot \size(C_{\calg}))$. 
\end{proof}

\subsection{Completing the proof}

We now have all the components necessary to complete the proof of \autoref{thm:explicit-map-ann-ub}.

\ExpMapAnnUB*

\begin{proof}
  Let $\calg' = \inparen{g_1(\vecz), \ldots, g_{2m}(\vecz)}$, the first $2m$ co-ordinates.
  Clearly, $C_{\calg}$ encodes $\calg'$ as well.

  \autoref{lem:ann-as-det} tells us that there is a $K \times K$ matrix, $\widetilde{M}$, such that $0 \not\equiv A'(\vecx) = \det (\widetilde{M})$ is an annihilator for the map $\calg'$ (that is, $A' \circ \calg' \equiv 0$).
  Here $K = {(2md+1)}^{2m}$ and $A'(\vecx)$ has individual degree at most $\ceil{{(2md)}^{m/(2m-m)}} + 1 = 2md + 1$.

  Using the fact that the entries of $\widetilde{M}$ are either evaluations of $\calg'$ or monomials, \autoref{lem:ann-as-succinct-det} provides a circuit $\widetilde{C}$ that encodes the matrix $\widetilde{M}$ where $\widetilde{C}$ has size $O((2md)^3 + (2m) \cdot \size(C_{\calg}))$.
  Further $\widetilde{C}$ is a purely algebraic circuit\footnote{Algebraic circuit without projection gates.} and also constant-free, assuming we have access to a gate computing $C_{\calg_m}$.
  Finally, we use \autoref{prop:succinct-determinant-vpspace} to obtain a circuit with projections $C'$ that computes $A'(\vecx)$.
  The circuit $C'$ has size $O((2md)^3 + (4m) \cdot \size(C_{\calg}) + \log^2 K) = O({(2md)}^3 + (4m) \cdot \size(C_{\calg})) = (m \cdot d \cdot s)^4$ for $s = \size(C_{\calg})$. 

  Finally, we note that for $A(x_1, \ldots, x_n) := A'(x_1, \ldots, x_{2m})$, $0 \not \equiv A(\vecx)$ and $A \circ \calg_m \equiv 0$.
  This completes the proof, since $C'$ also computes $A$.
\end{proof}

Clearly \autoref{thm:vp-ann-ub} is a special cases of \autoref{thm:explicit-map-ann-ub} and therefore follows as a corollary.
We restate it here for convenience.

\VPubAnn*

In fact, since the proof allows for the encoding circuit to use projection gates, we additionally have the following statement as a special case.

\begin{theorem}[Upper bound for annihilators of $\VPSPACE$]\label{thm:vpspace-ann-ub}
  For any $\VPSPACE_b$-explicit family of maps $\set{\calg_m}$, where each $\calg_m$ has more than $2m$ outputs, there is a family $\set{A_m}$ in $\VPSPACE_b$ such that $A_m$ annihilates $\calg_m$ for all large enough $m$.
\end{theorem}

\section{Lower Bounds from Hitting Set Generators}\label{sec:consequences}

We are now ready to formally prove our main results about the consequences of succinct and cryptographic hitting sets.
As mentioned before, these consequences follow by constructing annihilators for such hitting sets by applying \autoref{thm:explicit-map-ann-ub} with appropriate choices of parameters.

\subsection{Lower Bounds from Cryptographic Hitting Set Generators}\label{subsec:lb-from-crypto-hsg}

We begin by proving \autoref{thm:LBs-from-crypto-hsg}, which we first restate.

\CryptoHSGtoLB*

\begin{proof}
  We shall prove the statement via contradiction.
  Suppose that $\VP = \VNP$, and that uniform $\NC^1$ $\subseteq$ uniform $\TC^0$.
  Invoking the padding argument in \autoref{prop:TC0_NC1-vs-CH_SPACE}, we get that for any $m(n) = \poly(n)$, $\SPACE(m(n)) \subseteq \CH$, which means $\PSPACE \subseteq \CH$.  
  Thus, using \autoref{prop:vp-neq-vpspace-counting-hierarchy}, we get that $\VPSPACE_b = \VP$.

  This means that any family $\set{A_N}$ in $\VPSPACE_b$ of degree $d'(N) \in \poly(N)$ belongs to $\VP_{d'}$.
  For the specific parameters of the generator family $\set{\calh_n}$ in the hypothesis, \autoref{thm:explicit-map-ann-ub} implies a family $\set{A_N} \in \VPSPACE_b$ of annihilators of individual degree at most $(2n \cdot n^8) \leq N \cdot {(N/2)}^8 \leq N^{9}$, and thus total degree at most $N^{10} =: d(N)$.
  Now from our assumption, $\set{A_N} \in \VP_d$, which contradicts the HSG property of $\set{\calh_n}$.
  
  So it must be the case that either $\VP \neq \VNP$, or that uniform $\NC^1$ $\not\subseteq$ uniform $\TC^0$.
\end{proof}

One can weaken the hypothesis of \autoref{thm:LBs-from-crypto-hsg} to a family of $\VNP$ or even $\VPSPACE_b$-explicit HSGs because of \autoref{thm:vpspace-ann-ub}, as follows.
\begin{theoremwp}\label{thm:vpspace-hsg-to-vp-neq-vpspace}
  Let $\set{\calh_m}$ be a $\VPSPACE_b$-explicit family of $m$-variate polynomial maps with $n(m) > 2m$ outputs of degree $d_0(m) \in \poly(m)$, and let $d(n) \in \poly(n)$ be such that $d(n(m)) > 2m \cdot d_0(m)$, for all large $m$.
  
  If the family $\set{\calh_m}$ is a hitting set generator for $\VP_d$, then $\VP \neq \VPSPACE_b$.
  As a consequence, either $\VP \neq \VNP$ or $\mathsf{DLOGTIME}$-uniform $\NC^1$ $\not\subseteq$ $\mathsf{DLOGTIME}$-uniform $\TC^0$.
\end{theoremwp}

\begin{remark}[Known ``\textsf{PSPACE} constructions'']\label{rmk:about-supposed-pspace-constructions}
  Note that the $\PSPACE$-construction of hitting sets for algebraic circuits by Mulmuley~\cite{M12} (see also Forbes and Shpilka~\cite{FS18}), does not satisfy the hypothesis of \autoref{thm:vpspace-hsg-to-vp-neq-vpspace}.
  This is essentially the same reason as why the arguments of Heintz and Schnorr~\cite{HS80} do not give a ``cryptographic'' generator as required in \autoref{thm:LBs-from-crypto-hsg}.
  These constructions (\cite{M12,FS18}) yield a (possibly) \emph{different} family of generators constructible in space $\poly(n,d,s)$ for each size $s$.
  On the other hand, \autoref{thm:vpspace-hsg-to-vp-neq-vpspace} requires a single family that works for all sizes $s(n) \in \poly(n)$.
\end{remark}

\subsection{Upper Bound on Annihilators for Evaluation Vectors of $\mathsf{VP}$}\label{subsec:vp-ann-ub}

Next we prove \autoref{thm:vpspace-ub}.
Before moving on to the formal statement, we state some definitions and observations. 

\begin{definition}[Interpolating set]\label{defn:interpolating-set}
  For a set of polynomials $P$, a set of evaluation points $I$ is called an \emph{interpolating set for $P$}, if there is a \emph{linear transformation} $M$ such that for any polynomial $f \in P$, the coefficients of $f$ can be obtained from the evaluations of $f$ on the set $I$ by an application of $M$.
\end{definition}

\begin{propositionwp}[{Restatement of \cite[Theorem 10]{BP20}}]\label{prop:BP20-interpolating-set}
  Let $S := \set{0,1,2,\ldots,d}$, then the set of evaluation points $I_{n,d} := \set{ (a_1,a_2,\ldots,a_n) \in S^n \vert a_1+a_2+\cdots+a_n \leq d }$ forms an interpolating set for the set of all $n$-variate polynomials of total degree at most $d$.
\end{propositionwp}

\begin{definition}[Evaluation vector]\label{defn:evaluation-vector}
  For any $n$-variate polynomial $f(\vecx)$ of total degree-$d$, its evaluation vector is simply the vector\footnote{The order can be picked arbitrarily.} formed by evaluations of $f$ on the set $I_{n,d}$ from \autoref{prop:BP20-interpolating-set}.
\end{definition}

\begin{proposition}[Equations for evaluation vectors and natural proofs]\label{prop:evaluation-and-coefficient-vectors}
  Let $\calc$ and $\cald$ be either $\VP$ or $\VNP$.
  For any $d(n)$, there are $\cald$-natural proofs for $\calc_d$ if and only if there is a family $\set{A_N} \in \cald$ such that for each $\set{f_n} \in \calc_d$, $A_N$ vanishes on the evaluation vector of $f_n$ for all large enough $n$.
\end{proposition}
\begin{proof}[Proof Sketch]
  These statements essentially follow from the fact that for any $n$-variate, degree-$d$ polynomial, we can move between its coefficient vector and evaluation vector using linear transformations.
  Clearly, both these transformations have (constant-free) algebraic circuits of size $\poly(N)$.
  Since both $\VP$ and $\VNP$ are rich enough to implement the above circuits, we get all the required statements.
\end{proof}

We now state the upper bound on equations for \emph{evaluation vectors} of $\VP$.

\begin{lemma}\label{lem:main-vspace-ub}
  There exists a constant $c$, such that for all large enough $n,d,s \in \N$, a multilinear equation for the \emph{evaluation vectors} of the set of $n$-variate, degree-$d$ polynomials computable by circuits of size $s$, is computable by constant-free circuits with projection gates of size at most ${(nds)}^{c}$.
\end{lemma}
\begin{proof}
  For the given parameters $n,d,s$, let $N = \binom{n+d}{n}$ and consider the universal circuit $\calu(\vecx_{[n]},\vecy_{[r]})$ as given in \autoref{lem:size-of-universal-circuit}.
  Let $I \subseteq \C^n$ be an interpolating set of size $N$ given by \autoref{prop:BP20-interpolating-set}.
  Define a polynomial map $\calu_{n,d,s} : \C^r \rightarrow \C^N$ such that each co-ordinate of $\calu_{n,d,s}$ is exactly $\calu(\vecx=\veca,\vecy)$ for a unique point $\veca \in I$.
  We also know from \autoref{lem:size-of-universal-circuit} that $\calu_{n,d,s}$ can be encoded by a \emph{constant-free} circuit of size $\poly(n,d,s)$, and has individual degree at most $\poly(n,d,s)$.

  Since any $n$-variate, degree-$d$ polynomial that is computable by circuits of size $s$ can be obtained by fixing the $\vecy$ variables to some scalars in $\calu(\vecx,\vecy)$, we have that any annihilator of $\calu_{n,d,s}$ vanishes on the evaluation vector of \emph{every} such polynomial.
  
  Now, using the proof of \autoref{thm:explicit-map-ann-ub}, we can construct a \emph{multilinear}, $\poly(n,d,s)$-variate annihilator $A$ for $\calu_{n,d,s}$, that is computable by constant-free circuits with projections, of size at most ${(nds)}^{c}$ for some constant $c$.
  That the annihilator $A$ can be multilinear is guaranteed by the bound on the individual degree from \autoref{lem:ann-as-det}.
\end{proof}

\autoref{thm:vpspace-ub} then follows as a simple corollary. Observe that the $\log^{\ast}n$ in the exponent can be replaced by any growing function of $n$.

\VPSPACEub*

\begin{proof}
  Let $c$ be a constant as chosen in the proof of \autoref{lem:main-vspace-ub}, $\beta > 2$ be a constant to be fixed later and let $s = s(n) = n^{\log^{\ast}n / (\beta \cdot c)}$.
  Using \autoref{lem:main-vspace-ub}, define a multilinear polynomial family $\set{P_N}$ such that, for each $n$, $P_N(Z_1,\ldots,Z_N)$ is an annihilator for the set of $n$-variate, degree-$d$, size-$s$ polynomials and depends only on the first ${(nds)}^{2c} \leq n^{\log^{\ast}n} = t(n)$ variables.
  As each $P_N$ is computable using constant-free algebraic circuits with projections, of size $s^{3c} \leq t(n)$, the family $\set{P_N}$ is in $\VPSPACE^0$.
  Further, since the family of universal circuits $\set{\calu_{n,d,s}}$ is $\mathsf{DTIME}(\poly(n,d,s))$-uniform, using the characterization of $\VPSPACE^{0}$ due to Koiran and Perifel (\autoref{defn:VPSPACE0-coefficient}), the coefficient functions of $\set{P_N}$ can be computed by a Turing machine that uses space ${s(n)}^{3c \cdot \beta'}$ for some constant $\beta'$.
  We now fix $\beta = 3 \beta'$ so that ${s(n)}^{3c \cdot \beta'} \leq t(n)$.

  Let $\set{f_n} \in \VP_{d}$ be arbitrary.
  We know that for all large enough $n$, $\size(f_n) = n^a$ for some constant $a$.
  Thus, there is a finite $n_0 \in \N$ such that $\size(f_n) \leq s(n)$ for all $n > n_0$.
  Therefore, for all large enough $n$, the evaluation vectors of $f_n$ are zeroes of the polynomial $P_N(Z_1,\ldots,Z_N)$ and so $\set{P_N}$ is a family of equations whose coefficients are computable in (uniform) space $t(n)$.
\end{proof}

\subsection{Lower Bounds from Succinct Hitting Set Generators}\label{subsec:lb-from-succ-hsg}

Finally, we prove \autoref{thm:LBs-from-no-proofs} by combining the observations in the proofs of \autoref{thm:vpspace-ub} and \autoref{thm:LBs-from-crypto-hsg}.
We start by proving the following consequence of a separation between $\VPSPACE_b$ and $\VP$.
Here, by $\VPSPACE_b(m)$ we mean the class of $m$-variate polynomial families in $\VPSPACE_b$; when comparing this with $\VP(N)$, we treat them as $N$-variate families that depend only on the first $m$ inputs.
\begin{proposition}\label{prop:vp-neq-vpspace-final}
  Assuming the Generalized Riemann Hypothesis, suppose that for every growing function $f(n)$, and $m(n) := \log^{f(n)}n$, we have that $\VPSPACE_b(m) \not\subseteq \VP(n)$.

  Then either $\VP \neq \VNP$, or for every $g(N) = \omega(1)$ there is a boolean function family $\set{h_N}$ that depends only on its first $M = \exp((\log\log N)^{g(N)})$-inputs, such that:\\
  (a) $\set{h_N}$ has $\mathsf{DLOGTIME}$-uniform, $O(\log M)$-depth, $\poly(M)$-sized, fan-in 2 boolean circuits, and\\
  (b) $\set{h_N} \not\in $ $\mathsf{DLOGTIME}$-uniform $\TC^0$.\\
  Here the uniformity is in terms of the circuit size, in both cases.
\end{proposition}
\begin{proof}
  Firstly, note that the point (a) above is the same as saying: $\set{\tilde{h}_M} \in $ $\mathsf{DLOGTIME}$-uniform $\NC^1$, for the family $\set{\tilde{h}_M}$ defined so that $\tilde{h}_M = h_N$ for all $N$ large enough.
  We now prove the statement by contradiction.

  So suppose that the hypothesis holds, but not the conclusion.
  Therefore, $\VP = \VNP$ and for some $g(N) = \omega(1)$, every $\set{\tilde{h}_M}$ in $\mathsf{DLOGTIME}$-uniform $\NC^1$, has a $\mathsf{DLOGTIME}$-uniform constant-depth threshold circuit of size $N$, where $N$ is such that $M = \exp((\log\log N)^{g(N)})$.

  Now define $f(n) = g(2^n)$; note that $f(n) = \omega(1)$.
  Thus, from the hypothesis, we have that $\VPSPACE(\log^{f(n)}n) \not\subseteq \VP(n)$.
  Now using \autoref{prop:vp-neq-vpspace-counting-hierarchy}, we get that $\SPACE(\log^{f(n)}n) \not\subseteq \CH$.
  So, by instantiating \autoref{prop:TC0_NC1-vs-CH_SPACE} for $m(n) = \frac{\log^{f(n)}n}{a}$ (where $a$ is the absolute constant), we get that there is a function family $\set{p_L}$ on $L = 2^{\log^{f(n)}n}$ inputs, which has $\mathsf{DLOGTIME}$-uniform $\NC^1$ circuits, but requires $\mathsf{DLOGTIME}$-uniform constant depth threshold circuits of size $2^{n^{\omega(1)}}$.

  However, from our previous observation, taking $N = 2^{n}$, we have $L = 2^{m(n)} = 2^{(\log\log N)^{g(N)}}$, and therefore $\set{p_L}$ should have $\mathsf{DLOGTIME}$-uniform constant-depth threshold circuit of size $N$.
  This is a contradiction.
\end{proof}

\SuccHSGtoLB*
\begin{proof}
  Suppose we have $\VP_d$-succinct hitting sets for $\VP$, which means that for any $D(N),S(N) \in \poly(N)$, there is some family $\set{g^{(D,S)}_n} \in \VP_d$ that is a succinct hitting set generator for size-$S(N)$ families in $\VP_{D}$ (see \autoref{defn:succinct-hsg-vp-vnp}).
  Notice that over all choices of $D,S$, the family $\set{g^{(D,S)}_n}$ is a degree-$d(n)$ family; the different $\set{g^{(D,S)}_n}$s could potentially have different sizes $s(n) \in \poly(n)$.

  Now consider the universal circuit family $\set{\mathcal{U}_{n,d(n),t(n)}(\vecx,\vecy)}$ (\autoref{defn:universal-circuit-coeff-vector}) with a slightly super-polynomial size parameter, say $t(n) = n^{f(n)}$ for $f(n) = \omega(1)$.
  Observe that for every $\set{g^{(D,S)}_n} \in \VP_d$ above, there exists a large enough value of $n$, where $g^{(D,S)}_n(\vecx) = \mathcal{U}(\vecx,\vecy = \veca)$ for some $\veca$ in the field.
  As a result, the family $\set{\mathcal{U}_{n,d(n),t(n)}(\vecx,\vecy)}$ is a succinct hitting set generator for all of $\VP$.
  That is, for $N := \binom{n+d(n)}{d(n)}$, for any family $\set{P_N} \in \VP(N)$, there are infinitely many values of $N$ where $P_N$ will \emph{not vanish} on the coefficient vector of $\mathcal{U}$ (which is a vector of polynomials over the parameter variables $\vecy$).
  Using \autoref{prop:evaluation-and-coefficient-vectors}, this means that the same property holds even for the evaluation vectors of $\mathcal{U}$.
  Therefore, the family $\set{\mathcal{U}}$ is a family of degree-$d(n)$ polynomial maps that are encoded by circuits of size $t(n)$, such that any family of annihilators for these maps is outside $\VP(N)$.

  We can now apply \autoref{thm:explicit-map-ann-ub} to obtain a family of annihilators for $\set{\mathcal{U}}$, say $\set{A_m}$ in the class $\VPSPACE_b$, with $m = 2t(n)$, and degree at most $t(n)^5$ for all large enough $n$.
  Due to the hitting property of $\set{U}$, $\set{A_m} \not\in \VP(N)$, and thus, $\VPSPACE(t(n)) \not\subseteq \VP(N)$.
  Since this works for $t(n) = n^{f(n)}$ for any $f(n) = \omega(1)$, we are in the setting of \autoref{prop:vp-neq-vpspace-final}.

  Therefore, either $\VP \neq \VNP$, or for any $\gamma(N) = \omega(1)$ there is a family $\set{h_N}$ depending only on its first $M = \exp((\log\log N)^{\gamma(N)})$ inputs with the respective properties.
  We can then view this family as an $M$-variate family $\set{\tilde{h}_M}$ in uniform $\NC^1$ --- as in the proof of \autoref{prop:vp-neq-vpspace-final} --- that requires uniform constant-depth threshold circuits of size strictly larger than $N$.

  Finally, since $M$ is a growing function of $N$, there is some $\lambda(M) = \omega(1)$ such that $N \leq \exp(\exp(\log^{\sfrac{1}{\lambda(M)}}M))$, and we can then let $g(M) = \sfrac{1}{\lambda(M)} = o(1)$ to derive the second conclusion.
  Again, as the guarantee from \autoref{prop:vp-neq-vpspace-final} holds for any growing $\gamma(N)$, the conclusion also holds for any $g(M) = o(1)$ as required.

  \paragraph*{$\VP$-succinct hitting sets for $\VNP$.} In this case, we get that the universal circuit family $\set{U}$ is a hitting set generator for all of $\VNP$, using the exact same argument as above.
  As a result, the annihilator family $\set{A_m} \not\in \VNP$.
  Therefore, if it were the case that $\SPACE(\log^{\log^{\ast} N}N) \subseteq \P$, then the coefficient function of $\set{A_m}$ --- taking $m = n^{\sfrac{\log^{\ast}n}{c}} = \log^{\log^{\ast} N}N$ for some constant $c$ --- would be computable in $\operatorname{DTIME}(\poly(N))$, thus putting it in $\VNP$ using Valiant's criterion~\cite{V79a}, leading to a contradiction.
  This finishes the proof.
\end{proof}

\section*{Acknowledgements}

We thank C. Ramya and Ramprasad Saptharishi for discussions about this problem during its early stages.
We are grateful to Mrinal Kumar, Ramprasad Saptharishi and Amir Shpilka for numerous informative discussions on algebraic natural proofs, and algebraic complexity in general.

We thank Robert Andrews, Vishwas Bhargava, Christian Ikenmeyer, Meena Mahajan and Nitin Saxena for their helpful comments on a presentation of these results at the \href{https://sites.google.com/view/wac2023/home}{Worskhop on Algebra and Computation (WAC 2023)}; we also thank the organizers of this workshop for this opportunity.
Finally, we would like to acknowledge the workshop on \href{https://simons.berkeley.edu/workshops/proof-complexity-meta-mathematics}{Proof Complexity and Meta-mathematics} at the Simons institute, which facilitated various discussions regarding natural proofs with many experts in theoretical CS.
These interactions motivated this work and hence we thank the organizers of the workshop for inviting us.

We thank Robert Andrews for his comments on an earlier version of the paper that helped us improve the presentation.

\bibliographystyle{customurlbst/alphaurlpp}
\bibliography{masterbib/references,masterbib/crossref}

\newcommand{\etalchar}[1]{$^{#1}$}
\begin{thebibliography}{CMTV98}

\bibitem[AD08]{AD08}
Scott Aaronson and Andrew Drucker.
\newblock \href {https://www.scottaaronson.com/blog/?p=336} {Arithmetic natural proofs theory is sought}.
\newblock Shtetl Optimized: Scott Aaronson's Blog, 2008.

\bibitem[AV08]{AV08}
Manindra Agrawal and V.~Vinay.
\newblock \href {http://dx.doi.org/10.1109/FOCS.2008.32} {Arithmetic Circuits: A Chasm at Depth Four}.
\newblock In {\em \FOCS{2008}}, pages 67--75, 2008.
\newblock \shortECCC{08}{062}.

\bibitem[BDS24]{BDS24}
C.~S. Bhargav, Prateek Dwivedi, and Nitin Saxena.
\newblock \href {http://dx.doi.org/10.1007/978-981-97-2340-9\_30} {Lower Bounds for the Sum of Small-Size Algebraic Branching Programs}.
\newblock In {\em Theory and Applications of Models of Computation - 18th Annual Conference, {TAMC} 2024, Hong Kong, China, May 13-15, 2024, Proceedings}, volume 14637 of {\em Lecture Notes in Computer Science}, pages 355--366. Springer, 2024.

\bibitem[Ber84]{B84}
Stuart~J. Berkowitz.
\newblock \href {http://dx.doi.org/10.1016/0020-0190(84)90018-8} {On computing the determinant in small parallel time using a small number of processors}.
\newblock {\em Information Processing Letters}, 18(3):147 -- 150, 1984.

\bibitem[BP20]{BP20}
Markus Bl{\"{a}}ser and Anurag Pandey.
\newblock \href {http://dx.doi.org/10.4230/LIPIcs.APPROX/RANDOM.2020.8} {Polynomial Identity Testing for Low Degree Polynomials with Optimal Randomness}.
\newblock In {\em Approximation, Randomization, and Combinatorial Optimization. Algorithms and Techniques, {APPROX/RANDOM} 2020, August 17-19, 2020, Virtual Conference}, volume 176 of {\em LIPIcs}, pages 8:1--8:13. Schloss Dagstuhl - Leibniz-Zentrum f{\"{u}}r Informatik, 2020.

\bibitem[BS83]{BS83}
Walter Baur and Volker Strassen.
\newblock \href {http://dx.doi.org/10.1016/0304-3975(83)90110-X} {The Complexity of Partial Derivatives}.
\newblock {\em Theoretical Computer Science}, 22:317--330, 1983.

\bibitem[B{\"u}r00]{B00}
Peter B{\"u}rgisser.
\newblock \href {http://dx.doi.org/10.1007/978-3-662-04179-6} {{\em Completeness and Reduction in Algebraic Complexity Theory}}, volume~7 of {\em Algorithms and Computation in Mathematics}.
\newblock Springer, 2000.

\bibitem[B{\"{u}}r09]{B09}
Peter B{\"{u}}rgisser.
\newblock \href {http://dx.doi.org/10.1007/s00037-009-0260-x} {On Defining Integers And Proving Arithmetic Circuit Lower Bounds}.
\newblock {\em Computational Complexity}, 18(1):81--103, 2009.

\bibitem[CKR{\etalchar{+}}20]{CKRST20}
Prerona Chatterjee, Mrinal Kumar, C.~Ramya, Ramprasad Saptharishi, and Anamay Tengse.
\newblock \href {http://dx.doi.org/10.1109/FOCS46700.2020.00085} {On the Existence of Algebraically Natural Proofs}.
\newblock In {\em 61st {IEEE} Annual Symposium on Foundations of Computer Science, {FOCS} 2020, Durham, NC, USA, November 16-19, 2020}, pages 870--880. {IEEE}, 2020.
\newblock Pre-print available at \href {http://arxiv.org/abs/2004.14147} {\path{arXiv:2004.14147}}.

\bibitem[CKSS24]{CKSS24}
Prerona Chatterjee, Deepanshu Kush, Shubhangi Saraf, and Amir Shpilka.
\newblock \href {http://dx.doi.org/10.4230/LIPICS.CCC.2024.20} {Lower Bounds for Set-Multilinear Branching Programs}.
\newblock In {\em 39th Computational Complexity Conference, {CCC} 2024, July 22-25, 2024, Ann Arbor, MI, {USA}}, volume 300 of {\em LIPIcs}, pages 20:1--20:20. Schloss Dagstuhl - Leibniz-Zentrum f{\"{u}}r Informatik, 2024.

\bibitem[CKSV22]{CKSV22}
Prerona Chatterjee, Mrinal Kumar, Adrian She, and Ben~Lee Volk.
\newblock \href {http://dx.doi.org/10.1007/S00037-022-00223-8} {Quadratic Lower Bounds for Algebraic Branching Programs and Formulas}.
\newblock {\em Comput. Complex.}, 31(2):8, 2022.

\bibitem[CKW11]{CKW11}
Xi~Chen, Neeraj Kayal, and Avi Wigderson.
\newblock Partial Derivatives in Arithmetic Complexity.
\newblock {\em Foundations and Trends in Theoretical Computer Science}, 2011.

\bibitem[CMTV98]{CMTV98}
Herv{\'{e}} Caussinus, Pierre McKenzie, Denis Th{\'{e}}rien, and Heribert Vollmer.
\newblock \href {http://dx.doi.org/10.1006/JCSS.1998.1588} {Nondeterministic \emph{NC}\({}^{\mbox{1}}\) Computation}.
\newblock {\em J. Comput. Syst. Sci.}, 57(2):200--212, 1998.

\bibitem[Csa76]{C76}
L.~Csanky.
\newblock \href {http://dx.doi.org/10.1137/0205040} {Fast Parallel Matrix Inversion Algorithms}.
\newblock {\em {SIAM} J. Comput.}, 5(4):618--623, 1976.

\bibitem[FLMS15]{FLMS15}
Herv{\'{e}} Fournier, Nutan Limaye, Guillaume Malod, and Srikanth Srinivasan.
\newblock \href {http://dx.doi.org/10.1137/140990280} {Lower Bounds for Depth-4 Formulas Computing Iterated Matrix Multiplication}.
\newblock {\em SIAM Journal of Computing}, 44(5):1173--1201, 2015.
\newblock \pSTOC{2014}. \shortECCC{13}{100}.

\bibitem[FS12]{FS12}
Michael~A. Forbes and Amir Shpilka.
\newblock \href {http://dx.doi.org/10.1145/2213977.2213995} {On identity testing of tensors, low-rank recovery and compressed sensing}.
\newblock In {\em Proceedings of the 44th Symposium on Theory of Computing Conference, {STOC} 2012, New York, NY, USA, May 19 - 22, 2012}, pages 163--172. {ACM}, 2012.
\newblock Pre-print available at \parseECCC{TR11/147}.

\bibitem[FS18]{FS18}
Michael~A. Forbes and Amir Shpilka.
\newblock \href {http://dx.doi.org/10.1145/3188745.3188792} {A {PSPACE} construction of a hitting set for the closure of small algebraic circuits}.
\newblock In {\em \STOC{2018}}, pages 1180--1192. {ACM}, 2018.
\newblock \arXiv{1712.09967}.

\bibitem[FSV18]{FSV18}
Michael~A. Forbes, Amir Shpilka, and Ben~Lee Volk.
\newblock \href {http://dx.doi.org/10.4086/toc.2018.v014a018} {Succinct Hitting Sets and Barriers to Proving Lower Bounds for Algebraic Circuits}.
\newblock {\em Theory of Computing}, 14(1):1--45, 2018.
\newblock \pSTOC{2017}. \arXiv{1701.05328}.

\bibitem[GKKS14]{GKKS14}
Ankit Gupta, Pritish Kamath, Neeraj Kayal, and Ramprasad Saptharishi.
\newblock \href {http://dx.doi.org/10.1145/2629541} {Approaching the Chasm at Depth Four}.
\newblock {\em Journal of the ACM}, 61(6):33:1--33:16, 2014.
\newblock \pCCC{2013}.
\newblock Pre-print available at \parseECCC{TR12/098}.

\bibitem[GKSS17]{GKSS17}
Joshua~A. Grochow, Mrinal Kumar, Michael~E. Saks, and Shubhangi Saraf.
\newblock \href {http://arxiv.org/abs/1701.01717} {Towards an algebraic natural proofs barrier via polynomial identity testing}.
\newblock {\em CoRR}, abs/1701.01717, 2017.
\newblock Pre-print available at \href {http://arxiv.org/abs/1701.01717} {\path{arXiv:1701.01717}}.

\bibitem[GR05]{GR05}
Ariel Gabizon and Ran Raz.
\newblock Deterministic Extractors for Affine Sources over Large Fields.
\newblock In {\em \FOCS{2005}}, pages 407--418, 2005.

\bibitem[Gro15]{G15}
Joshua~A. Grochow.
\newblock \href {http://dx.doi.org/10.1007/s00037-015-0103-x} {Unifying Known Lower Bounds via Geometric Complexity Theory}.
\newblock {\em Computational Complexity}, 24(2):393--475, 2015.

\bibitem[HILL99]{HILL99}
Johan H{\aa}stad, Russell Impagliazzo, Leonid~A. Levin, and Michael Luby.
\newblock \href {http://dx.doi.org/10.1137/S0097539793244708} {A Pseudorandom Generator from any One-way Function}.
\newblock {\em {SIAM} J. Comput.}, 28(4):1364--1396, 1999.

\bibitem[HS80]{HS80}
Joos Heintz and Claus-Peter Schnorr.
\newblock \href {http://dx.doi.org/10.1145/800141.804674} {Testing Polynomials which Are Easy to Compute (Extended Abstract)}.
\newblock In {\em \STOC{1980}}, pages 262--272, 1980.

\bibitem[Kal85]{K85}
Kyriakos Kalorkoti.
\newblock \href {http://dx.doi.org/10.1137/0214050} {{A Lower Bound for the Formula Size of Rational Functions}}.
\newblock {\em SIAM Journal of Computing}, 14(3):678--687, 1985.

\bibitem[Kay09]{K09}
Neeraj Kayal.
\newblock \href {http://dx.doi.org/10.1109/CCC.2009.37} {The Complexity of the Annihilating Polynomial}.
\newblock In {\em Proceedings of the 24th Annual {IEEE} Conference on Computational Complexity, {CCC} 2009, Paris, France, 15-18 July 2009}, pages 184--193, 2009.

\bibitem[KI04]{KI04}
Valentine Kabanets and Russell Impagliazzo.
\newblock \href {http://dx.doi.org/10.1007/s00037-004-0182-6} {{D}erandomizing Polynomial Identity Tests Means Proving Circuit Lower Bounds}.
\newblock {\em Computational Complexity}, 13(1-2):1--46, 2004.
\newblock \pSTOC{2003}.

\bibitem[KP09]{KP09}
Pascal Koiran and Sylvain Perifel.
\newblock \href {http://dx.doi.org/10.1007/s00037-009-0269-1} {{VPSPACE} and a Transfer Theorem over the Reals}.
\newblock {\em Computational Complexity}, 18(4):551--575, 2009.

\bibitem[KRST22]{KRST22}
\mfbiberr{toupdate(KRST22): journal}Mrinal Kumar, C.~Ramya, Ramprasad Saptharishi, and Anamay Tengse.
\newblock \href {http://dx.doi.org/10.4230/LIPIcs.STACS.2022.44} {If {VNP} Is Hard, Then so Are Equations for It}.
\newblock In {\em 39th International Symposium on Theoretical Aspects of Computer Science, {STACS} 2022, March 15-18, 2022, Marseille, France (Virtual Conference)}, volume 219 of {\em LIPIcs}, pages 44:1--44:13. Schloss Dagstuhl - Leibniz-Zentrum f{\"{u}}r Informatik, 2022.
\newblock \arXiv{2012.07056}.

\bibitem[Mal11]{M11}
Guillaume Malod.
\newblock \href {http://dx.doi.org/10.1007/978-3-642-22953-4\_18} {Succinct Algebraic Branching Programs Characterizing Non-uniform Complexity Classes}.
\newblock In {\em Fundamentals of Computation Theory - 18th International Symposium, {FCT} 2011, Oslo, Norway, August 22-25, 2011. Proceedings}, pages 205--216, 2011.

\bibitem[MR13]{MR13}
Meena Mahajan and B.~V.~Raghavendra Rao.
\newblock \href {http://dx.doi.org/10.1007/s00037-011-0024-2} {Small Space Analogues of Valiant's Classes and the Limitations of Skew Formulas}.
\newblock {\em Computational Complexity}, 22(1):1--38, 2013.

\bibitem[Mul12]{M12}
Ketan Mulmuley.
\newblock \href {http://dx.doi.org/10.1109/FOCS.2012.15} {Geometric Complexity Theory {V:} Equivalence between Blackbox Derandomization of Polynomial Identity Testing and Derandomization of Noether's Normalization Lemma}.
\newblock In {\em \FOCS{2012}}, pages 629--638, 2012.

\bibitem[MV97]{MV97}
Meena Mahajan and V.~Vinay.
\newblock \href {http://dx.doi.org/10.1.1.31.1673} {A Combinatorial Algorithm for the Determinant}.
\newblock In {\em \SODA{1997}}, pages 730--738, 1997.
\newblock Available on \href{http://citeseerx.ist.psu.edu/viewdoc/summary?doi=10.1.1.31.1673}{\tt citeseer:10.1.1.31.1673}.

\bibitem[Poi08]{P08b}
Bruno Poizat.
\newblock \href {http://dx.doi.org/10.2178/jsl/1230396913} {A la recherche de la definition de la complexite d'espace pour le calcul des polynomes a la maniere de Valiant}.
\newblock {\em J. Symb. Log.}, 73(4):1179--1201, 2008.

\bibitem[Raz10]{R10a}
Ran Raz.
\newblock \href {http://dx.doi.org/10.4086/toc.2010.v006a007} {Elusive Functions and Lower Bounds for Arithmetic Circuits}.
\newblock {\em Theory of Computing}, 6(1):135--177, 2010.

\bibitem[RR97]{RR97}
Alexander~A. Razborov and Steven Rudich.
\newblock \href {http://dx.doi.org/10.1006/jcss.1997.1494} {Natural Proofs}.
\newblock {\em Journal of Computer and System Sciences}, 55(1):24--35, 1997.
\newblock \pSTOC{1994}.

\bibitem[Sap15]{S15}
Ramprasad Saptharishi.
\newblock \href {https://github.com/dasarpmar/lowerbounds-survey/releases/} {A survey of lower bounds in arithmetic circuit complexity}.
\newblock Github survey, 2015.

\bibitem[Smo97]{S97b}
Roman Smolensky.
\newblock \href {http://dx.doi.org/10.1007/BF01294255} {Easy Lower Bound for a Strange Computational Model}.
\newblock {\em Comput. Complex.}, 6(3):213--216, 1997.

\bibitem[Str73]{S73}
Volker Strassen.
\newblock \href {http://dx.doi.org/10.1007/BF01436566} {Die {B}erechnungskomplexit\"{a}t Von Elementarsymmetrischen {F}unktionen Und Von {I}nterpolationskoeffizienten}.
\newblock {\em Numerische Mathematik}, 20(3):238--251, 1973.

\bibitem[SY10]{SY10}
Amir Shpilka and Amir Yehudayoff.
\newblock \href {http://dx.doi.org/http://dx.doi.org/10.1561/0400000039} {Arithmetic Circuits: A survey of recent results and open questions}.
\newblock {\em Foundations and Trends in Theoretical Computer Science}, 5:207--388, March 2010.

\bibitem[Val79a]{V79a}
Leslie~G. Valiant.
\newblock \href {http://dx.doi.org/10.1145/800135.804419} {{Completeness Classes in Algebra}}.
\newblock In {\em \STOC{1979}}, pages 249--261, 1979.

\bibitem[Val79b]{V79b}
Leslie~G. Valiant.
\newblock \href {http://dx.doi.org/10.1016/0304-3975(79)90044-6} {The Complexity of Computing the Permanent}.
\newblock {\em Theor. Comput. Sci.}, 8:189--201, 1979.

\end{thebibliography}

\appendix

\section{Equivalence of the definitions of \textsf{VPSPACE}}\label{sec:vpspace-equiv-defns}

This section is devoted to proving \autoref{prop:equivalence-VPSPACE-definitions}, restated below.

\VPSPACEequivalence*

Throughout this section, whenever a bit-string corresponds to an integer, the $0^{th}$ bit refers to the sign-bit, the first bit to the most significant bit (MSB) and the last bit to the least significant bit (LSB).

Before moving to \autoref{prop:equivalence-VPSPACE-definitions}, we sketch the proofs of a few basic facts that will be needed.

\begin{observation}[Adding in small space]~\label{obs:add-log-space}
  Let $A, B$ be positive integers such that $A+B$ is at most $2^{2^{s}}$.
  Assuming bit-access to $A$ and $B$, any bit of the sum $A+B$ can be computed in space at most $O(s)$.
\end{observation}
\begin{proof}[Sketch]
  Let $i \in [s]$ be the index of the bit of $A+B$ that we wish to compute.
  This will be done in $i$ iterations, starting from the LSB to the $i^{th}$ bit.
  We maintain a counter $j$ (using $s$ bits) to keep track of the current iteration.
  We also use $4$ additional bits: $a$, $b$, $r$ and $c$, for $A$, $B$, the result and the carry, respectively, all initialized to $0$.
  
  In iteration $j$, we get the $j^{th}$ bits of $A$ and $B$ in $a$ and $b$, respectively.
  We then set $r \leftarrow a \oplus b \oplus c$, and $c \leftarrow (a \wedge b) \vee (b \wedge c) \vee (a \wedge c)$, in that order.
  At the end of iteration $i$, we output $r$ as the $i^{th}$ bit of $A+B$.

  The total space used is, clearly, at most $O(s)$ for all large $s$.
\end{proof}

\begin{observation}[Subtracting in small space]~\label{obs:subtract-log-space}
  Let $A, B$ be integers of absolute value at most $2^{2^{s}}$ such that the absolute value of $A-B$ is at most $2^{2^{s}}$.
  Assuming bit-access to $A$ and $B$, any bit of the difference $A - B$ can be computed in space at most $O(s)$.
\end{observation}
\begin{proof}[Sketch]
  As an initial step, we query the sign-bits ($0^{th}$ bits) of $A$ and $B$, to decide whether the absolute value of $A-B$ is the sum of the difference of their absolute values.

  The rest of the algorithm is the same as that for \autoref{obs:add-log-space}, but with the correct expressions for computing $r$ and $c$ in each of the iterations.
  In order to compute the sign-bit of $A - B$, we need to check the sign-bits of $A$ and $B$, and check if there is a carry from iteration $1$.
\end{proof}

\begin{observation}[Summing a list in small space]~\label{obs:list-add-log-space}
  Let $A_1, A_2, \ldots, A_{L}$ be positive integers such that their sum $T := A_1 + A_2 + \cdots + A_L$ is at most $2^{2^s}$.
  Assuming bit-access to each of the $A_i$s, any bit of the sum $T$ can be computed in space at most $O(\ceil{\log L} \cdot s)$.
\end{observation}
\begin{proof}[Sketch]
  We prove this using \autoref{obs:add-log-space}, and induction on the size of the list, $L$.
  Assuming bit-access to the sums $T_1 = A_1 + \cdots + A_{L/2}$, and $T_2 = A_{L/2 + 1} + \cdots + A_{L}$, we can compute the sum $T$ in space $2s$.
  In order to simulate bit-access to $T_1$ and $T_2$, we recursively run the algorithm of the two smaller lists.
  The key observation is that we can reuse the same $2 \cdot (\ceil{\log L} - 1) \cdot s$ bits of space for both these computations.
  The total space used is thus, at most $O(\ceil{\log L} \cdot s)$.
\end{proof}

\begin{observation}[Multiplying in small space]~\label{obs:multiply-log-space}
  Let $A, B$ be positive integers such that $A \times B$ is at most $2^{2^s}$.
  Assuming bit-access to $A$ and $B$, any bit of the product $A \times B$ can be computed in space at most $O(s^3)$.
\end{observation}
\begin{proof}[Sketch]
  Using the ``school method'' the product $A \times B$ can be calculated as the sum of $L = \ceil{\log B}$ many numbers, each of which is either $0$, or a shift of $A$.

  Therefore, we can use \autoref{obs:list-add-log-space} by simulating bit-access to each of the shifts mentioned above.
  Here, the $j^{th}$ bit in the $k^{th}$ shift of $A$, is either $0$ (if the $k^{th}$ bit of $B$ is $0$), or the ${(j+k-1)}^{th}$ bit of $A$.
  So we just need an additional space of at most $2s$ bits for this simulation.
  
  The total space used is therefore $O(s^2) + O(s) \leq O(s^3)$, for all large enough $s$.
\end{proof}

We are now ready to prove \autoref{prop:equivalence-VPSPACE-definitions}, which we restate once more.

\VPSPACEequivalence*

The statement clearly follows from the following lemmas, namely \autoref{lem:projections-to-coefficients} and \autoref{lem:coefficients-to-projections}.

\begin{lemma}~\label{lem:projections-to-coefficients}
  Let $s$ be large enough, and suppose $f(\vecx) \in \Z[\vecx]$ is computable by a constant-free, algebraic circuit with projections gates, of size $s$.

  Then, the coefficient function\footnote{\autoref{defn:coefficient-function}.} $\Phi$ of $f$, is computable by a Turing machine that uses space at most $O(s^4)$, and takes as advice a string of length at most $O(s^2)$.
\end{lemma}
\begin{proof}
  Let $C$ be the constant-free circuit with projection gates that computes $f$.
  The advice to the Turing machine is going to be an encoding of the graph of $C$.

  We can assume that $C(\vecx) = C_1(\vecx) - C_2(\vecx)$, with $C_1,C_2$ being monotone (that is, they only use the positive constant $1$, and addition and multiplication gates).
  We shall therefore assume that $C$ itself is monotone, and show that its coefficient function can be computed in small space, and then just use \autoref{obs:subtract-log-space} to finish the proof.

  For each gate $g \in C$, let $\Phi_g$ be its coefficient function.
  \begin{itemize}\itemsep0em
    \item If $g$ is a leaf, then its coefficient function is trivial.
    
    \item If $g = u + v$, then $\Phi_g$ can be decided in space $O(s)$ using \autoref{obs:add-log-space}, by using an additional space of at most $O({(s-1)}^4)$ bits, to simulate $\Phi_u$ and $\Phi_v$ whenever required.
    
    \item If $g = \proj_{z=0} u(z,\vecx)$, then $\Phi_g$ is essentially $\Phi_u$ with the exponent for $z$ fixed to zero.\\
    When $g = \proj_{z=1} u(z,\vecx)$, for any monomial $\vecx^{\vece}$, $\coeff_g(\vecx^{\vece}) = \sum_{0 \leq a \leq \deg_z(u)} \coeff_{u}(\vecx^{\vece} \cdot z^a)$.
    Therefore, we can calculate $\Phi_g$ using $\Phi_u$ and \autoref{obs:list-add-log-space}.
    
    \item If $g = u \times v$, then the coefficient of a monomial $\vecx^{\vece}$ in $g$ is given by $\sum_{\vece'} \coeff_u(\vecx^{\vece'}) \coeff_v(\vecx^{\vece - \vece'})$.
    The degree of $f$ is at most $2^s$, and hence there are at most $2^{ns} \leq 2^{s^2}$ terms in the above sum.
    Therefore, if we can simulate bit-access to each of the product terms, then we can compute any bit of the sum in additional space $O(s^3)$ using \autoref{obs:list-add-log-space}.
    Since we can inductively compute the coefficient functions of both $u$ and $v$in space $O({(s-1)}^4)$, \autoref{obs:multiply-log-space} gives us bit-access to each of the terms in the sum, using an additional space of at most $O(s^3)$ bits.
    Putting all the pieces together, $\Phi_g$ can be decided in space at most $O({(s-1)}^4 + s^3)$.
  \end{itemize}
  
  Thus, the total space used is at most $O({(s-1)}^4 + s^3) \leq O(s^4)$, for all large $s$.
\end{proof}

\begin{lemma}\label{lem:coefficients-to-projections}
  Let $f(\vecx) \in \Z[\vecx]$ be an $n$-variate, degree $d$ polynomial such that its coefficient function can be computed in space $s$.
  Then there is a constant-free circuit with projection gates of size $\poly(s, n, \log d)$ that can compute $f(\vecx)$. 
\end{lemma}

\begin{proof}
  We prove this in two steps.
  First we show that the coefficient function can be efficiently computed by a constant-free algebraic circuit with projection gates.
  We then show that if the coefficient function of a polynomial can be efficiently computed by a constant-free algebraic circuit with projection gates, then the polynomial itself can be efficiently computed by a constant-free algebraic circuit with projection gates.

  \begin{claim}
    Let $f(\vecx) \in \Z[\vecx]$ be an $n$-variate, degree $d$ polynomial such that its coefficient function can be computed in space $s$.
    Then there is a constant-free algebraic circuit with projection gates of size $\poly(s,n,\log d)$ that computes the coefficient function of $f$.
  \end{claim}

  \begin{proof}
    Using the fact that totally quantified boolean expressions capture $\PSPACE$, since the coefficient function of $f$ can be computed using space $s$, we note that the coefficient function can be written as a quantified boolean formula of size $\poly(s)$.

    We now \emph{arithmetize} the boolean expression in the usual way to get an algebraic circuit.
    Additionally, we use projection gates to simulate the quantifiers using constant sized gadgets: 
    \begin{itemize}
      \item $\forall y \Psi(y) \equiv \proj_{y=0} p(y) \times \proj_{y=1} p(y)$,
      \item $\exists y \Psi(y) \equiv 1 - (1 - \proj_{y=0} p(y)) \times (1 - \proj_{y=1} p(y))$.
    \end{itemize}
    We therefore get an algebraic circuit with projection gates, of size $\poly(s,n,\log d)$ that computes the coefficient function of $f$.
  \end{proof}

  \begin{claim}
    Let $f(\vecx) \in \Z[\vecx]$ be an $n$-variate, degree $d$ polynomial such that its coefficient function can be computed by a constant-free algebraic circuit with projection gates of size $s'$.
    Then there is a constant-free algebraic circuit with projection gates of size $\poly(s',n,\log d)$ that computes $f$.
  \end{claim}

  \begin{proof}
    It is easy to see that projection gates can efficiently simulate exponential sums.
    So it suffices to provide an exponential sum that computes $f(\vecx)$ given access to a circuit, say $\mathrm{CF}(\vecy_e,\veci)$, computing the coefficient function, as follows.
    Here $m = n \cdot \ceil{\log d}$ is the length of the bit-vectors for exponents, and $b$ is the base-$2$-logarithm of the bit complexity of the coefficients of $f$.
    \[
      f(\vecx) = \sum_{\vecy_{e} \in \set{0,1}^m} \inparen{\sum_{\veci \in \set{0,1}^b} \mathrm{pow}(\veci) \cdot \mathrm{CF}(\vecy_e,\veci) } \cdot \mathrm{check}(\vecy_e) \cdot \mathrm{mon}(\vecx,\vecy_e)
    \]
    The polynomials $\mathrm{pow}$ computes the $i^{th}$ power of two, $\mathrm{check}$ outputs $1$ when the exponent $\vece$ is valid, and zero otherwise, and $\mathrm{mon}$ computes the $\vece^{th}$ monomial in $\vecx$.
    All these polynomials have easy constructions in size $\poly(m,b) = \poly(n,\log d, s)$.
  \end{proof}
  
  The required statement clearly follows from the above two claims.
\end{proof}

\end{document}